\documentclass[aip,floatfix,amsmath,amssymb,reprint,superscriptaddress]{revtex4-2}

\usepackage{natbib}

\usepackage{siunitx}
\usepackage{amsmath}
\usepackage{svg}
\usepackage{amsmath} 
\usepackage[version=4,arrows=pgf-filled,
textfontname=sffamily,
mathfontname=mathsf]{mhchem}

\usepackage{hyphenat}
\usepackage{graphicx}
\usepackage{microtype}

\begin{document}

\hyphenchar\font=\string"7F

\let\WriteBookmarks\relax
\def\floatpagepagefraction{1}
\def\textpagefraction{.001}

\title{Neutron Reflectometry Reveals Conformational Changes in a Mechanosensitive  Protein Induced by an Antimicrobial Peptide in Tethered Lipid Bilayers}

%%%%%%%%%%%%%%%%%%%%%%%%%
\author{Sophie E. Ayscough}
\affiliation{School of Physics \& Astronomy, James Clerk Maxwell Building, University of Edinburgh, Edinburgh, EH9 3FD, UK}
\affiliation{Lund University, Lund 22100, Sweden}
\author{Maximilian W. A. Skoda}
\email{maximilian.skoda@stfc.ac.uk}
\affiliation{ISIS Neutron and Muon Source, Rutherford Appleton Laboratory, Harwell Campus, Chilton, OX11 0QX, UK}
\author{James Doutch}
\affiliation{ISIS Neutron and Muon Source, Rutherford Appleton Laboratory, Harwell Campus, Chilton, OX11 0QX, UK}
\author{Andrew Caruana}
\affiliation{ISIS Neutron and Muon Source, Rutherford Appleton Laboratory, Harwell Campus, Chilton, OX11 0QX, UK}
\author{Christy Kinane}
\affiliation{ISIS Neutron and Muon Source, Rutherford Appleton Laboratory, Harwell Campus, Chilton, OX11 0QX, UK}
\author{Luke Clifton}
\affiliation{ISIS Neutron and Muon Source, Rutherford Appleton Laboratory, Harwell Campus, Chilton, OX11 0QX, UK}
\author{Simon Titmuss}
\email{simon.titmuss@ed.ac.uk}
\affiliation{School of Physics \& Astronomy, James Clerk Maxwell Building, University of Edinburgh, Edinburgh, EH9 3FD, UK}

\begin{abstract}
\textit{Hypothesis} Membrane proteins serve a wide range of vital roles in the functioning of living organisms. They are responsible for many cellular functions, such as signaling, ion and molecule transport, binding and catalytic reactions. Compared to other classes of proteins, determining membrane protein structures remains a challenge, in large part due to the difficulty in establishing experimental conditions that can preserve the correct conformation and function of the protein in isolation from its native environment. Many therapeutics target membrane proteins which are accessible on the surface of cells. Here we hypothesize that the observed efficacy of antimicrobial peptides (AMPs) that interact with bacterial membranes may in part be associated with their triggering of a conformational change in the Mechansensitive Ion Channel of Large Conductance (MscL). 

\textit{Experiments}  
We investigated the ion channel in lipid vesicles and in a planar lipid bilayer. We developed a novel method for protein-lipid planar bilayer formation, avoiding the use of detergents. By using a polymeric tether our planar membrane mimetic was not constrained by the underlying solid substrate, making it sufficiently flexible to allow for increases in bilayer curvature and changes in membrane tension. We used quartz crystal microbalance  with dissipation (QCM-D), and polarised neutron reflectivity (PNR) to show the formation of MscL containing phospholipid bilayers, tethered with a high density PEG layer onto gold substrates from vesicle rupture.  The MscL containing vesicles were separately characterised with small angle neutron scattering (SANS).

\textit{Findings} MscL was expressed into vesicles using cell free protein expression. Analysing these vesicles with small angle neutron scattering, the radius of gyration of the protein was determined to be between 26-29~\AA{}, consistent with the crystal structure of individual MscL channels. The MscL composition of the formed bilayer was 14\%$v/v$, close to the initial composition of the vesicles, and a protein protrusion extending ca. 46~\AA{} into the solvent was determined by PNR. Addition of 1.6 and 3.2 $\mu$M pexiganan resulted in a decrease in the protrusion of MscL (from $\sim$46 to $\sim$38~\AA{}). To our knowledge, these findings represent the first direct experimental evidence of a structural change in the C-terminus containing protrusion of MscL, triggered by an antimicrobial peptide. 
\end{abstract}

%        \includegraphics[width=\linewidth]{FigureGA.pdf}

%\item We present a method for the formation of a lipid bilayer containing mechanosensitive ion channels that avoids the use of detergent.
%\item Small angle neutron scattering from the ion channel containing vesicles  indicates the ion channels are expressed as individual channels rather than clusters.
%\item The increase in radius of gyration and the decrease in the extent of the C-terminus containing protrusion determined by SANS and polarised neutron reflectometry, respectively, that result from the addition of the antimicrobial peptide, pexiganan, could be consistent with the ion channel opening.

% \phantomsection
% Keywords
%\begin{keywords}
% mechanosensitive ion channel \sep MscL \sep tethered bilayer \sep neutron reflectivity \sep small-angle neutron scattering
%\end{keywords}

\maketitle

% Main text


%%%%%%%%%%%%%%%%%%%%%%%%%%%%
\section{Introduction}
%%%%%%%%%%%%%%%%%%%%%%%%%%%%
Antimicrobial resistance (AMR) is a pressing worldwide health challenge, leading to complications in treating bacterial infections. Recognized as one of the top ten global public health threats by the World Health Organization (WHO), some studies suggest that AMR is anticipated to cause approximately 10 million fatalities annually by the year 2050.\cite{ ONeill2016}
To address this issue, numerous alternative therapies have been proposed, and among them, antimicrobial peptides (AMPs) emerged as a highly promising option over two decades ago. These peptides have been naturally present for millions of years with minimal to no development of resistance\cite{Yu2018}, making them an appealing alternative to antibiotics, to which bacteria tend to develop resistance rapidly. The effectiveness of AMPs in countering microbial resistance is attributed to their diverse modes of action against bacteria compared to the fixed targets of antibiotics.\cite{Nicolas2009}
Furthermore, AMPs are considered less toxic since they are broken down into amino acids, unlike other treatments that might produce potentially harmful metabolites. This unique characteristic adds to their attractiveness as potential therapeutic agents. It is important to note however, that bacteria have been shown to become resistant to certain AMPs,\cite{assoni2020resistance, gruenheid2012resistance} thus increasing our understanding of their mechanism could result in development and identification of drugs with low resistance rates.
Research into AMPs has mostly focused on peptide discovery and characterisation of the mechanism by which peptides target and destabilise membranes. Current research suggests a variety of ways in which AMPs could form pores in the membrane leading to membrane lysis. Popular models are the toroidal pore model,\cite{ludtke1995membrane} the barrel stave model\cite{He1996} and the carpet model\cite{Fernandez2012} -- the last of which describes destabilisation of the membrane without the need for pore formation. Pexiganan is thought to insert at the interface between head group and the acyl tail regions.\cite{gottler2009structure}

Regarding specificity, it is the predominantly anionic lipid bilayer of bacterial membranes which is targeted by cationic AMPs.\cite{Gottler2009} The role of charge in the interaction confers selectivity towards bacteria rather than eukaryotic membranes, which contain a lower fraction of anionic lipids, thus reducing toxicity. Given the relative clinical success of pexiganan compared to other antimicrobial peptides, it has been widely used as an archetype for short-chained cationic AMPs and hence we decided to use it as a model peptide in this work.\cite{gottler2009structure}

The MscL (Mechanosensitive Ion Channel of Large Conductance) is a highly conserved membrane protein in prokaryotes, that acts as an emergency pressure release valve in response to the cell being subjected to osmotic shock.\cite{levina1999protection, sukharev1994large} It is one of several mechanosensitive ion channels found in prokaryotes, but since it forms the largest pore and is the most conserved across several species it is the most attractive drug target. \cite{wang2023feeling}  Indeed one study named it as one of the top 20 potential drug targets.\cite{barh2011novel} When the channel is open the membrane is permeabilised temporarily to restore osmotic pressure to the cell, allowing for influx and efflux of compounds and osmolytes. It has been shown that inappropriate gating of MscL channels is highly deterimental to the cells. \cite{ou1998one} Given that therapeutic gating of the channels may be possible through direct or indirect drug interactions, MscL is a viable and exciting drug target, the understanding of which could alleviate the antiobiotic-resistance crisis.

Various studies have proposed mechanosensitive channels as determinants of antibacterial susceptibility. \cite{WANG2023, Kouwen2009, Boyan2022} Regarding the mode of action,  some studies suggest that they serve as entry gates for antimicrobials into cells, thus enhancing antibiotic efficiency, while others propose that they play a role in antibiotic-stress adaptation, reducing susceptibility to certain antimicrobials. Kouwen et al.\cite{Kouwen2009} on the other hand, have identified the critical role of the MscL channel (Mechanosensitive Ion Channel of Large Conductance) in the susceptibility of \textit{Bacillus subtilis} and \textit{Staphylococcus aureus} toward the lantibiotic (small post-translationally modified peptides with antimicrobial activity) sublancin 168. It was shown that MscL may serve either as a direct target for this lantibiotic or as a gate of entry to the cytoplasm. Wray et al. also identified MscL gating as the mechanism by which the antibiotic dihydrostreptomycin (DHS) enters the cytoplasm of the cells. \cite{wray2016dihydrostreptomycin} 

The crystal structure of the MscL homolog from Mycobacterium tuberculosis was already determined in 1998.\cite{chang1998structure, wray2016dihydrostreptomycin} As a transmembrane protein it spans  both the lipid head group and tail regions of the bilayer. It is also expected to protrude from the membrane by about 40~\AA{} (C-terminal end) on the basis of the crystal structure listed as 2OAR on the protein structure database.~\cite{Rees2007} Molecular dynamics simulations paint a somewhat different picture, suggesting a more disordered structure of the C-terminal protrusion.\cite{GULLINGSRUD2001, GULLINGSRUD2003}
Knowledge of the crystal structure of the closed channel aided research groups to make further studies into the function and behaviour of the mechanosensitve ion channel.\cite{Martinac2004, perozo2002open} MscL is known to form a water-filled pore in the membrane of 30~\AA{} diameter in response to osmotic shock, a change in potential across the membrane and in response to the insertion of amphiphillic molecules into the membrane.\cite{martinac1990mechanosensitive, moe1998functional} There is consensus in the literature that bacterial mechanosensitive channels directly sense membrane tension developed solely in the lipid bilayer - a mechanism named the bilayer mechanism (refs.~\cite{Corry2008, Phillips09} and refs therein). This mechanism is thought to occur in two cases: (i) protein–lipid-bilayer hydrophobic mismatch (when bilayer thinning leads to exposure of hydrophobic regions in MscL) and (ii) bilayer curvature, which can be induced by insertion of an amphipathic molecule into the bilayer.

By examining the structural response of MscL containing bilayers to challenge by pexiganan, this study aims to provide insight into whether MscL could be a target for AMPs. Following the original observation of Martinac {\it et al.}  that the presence of amphipathic drug molecules increased the probability of the MscL channel being open,\cite{martinac1990mechanosensitive} we consider the possibility that the insertion of the amphipathic AMP, pexiganan, alters the membrane tension/curvature sufficiently to provoke a permanent gating open of MscL, similar to that caused by lyso-PC.
Facilitating uncontrolled efflux of material from the cell through an unselective pore would present a method by which peptides could cause bacterial cell death at a lower concentration then at which formation of toroidal pores or membrane solubilisation have been shown to occur. \cite{hallock2003msi, gong2023antimicrobial} Alternatively, permanent opening of the channel would allow for ingress of antibiotic molecules, such as streptomycin,\cite{wray2016dihydrostreptomycin} used as part of a combination therapy with AMPs.

Lattice-like clustering of membrane proteins has been observed in several systems and may be a mechanism by which bacteria and other cells modulate protein function. \cite{engelman2005membranes} It has been shown through Atomic Force microscopy measurements (AFM) that when highly expressed in model membranes, MscL can form clusters and patch clamp measurements  further indicated protein-protein interaction and modulation of gating sensitivity.\cite{grage2011bilayer} It is therefore of interest to attempt to identify any evidence of clustering in our model system.

In this paper, we use small angle neutron scattering to demonstrate that MscL was succesfully expressed by cell free protein expression (CFPE) into mixed POPC:POPG (3:1) vesicles, and the interaction of the antimicrobials pexiganan and lyso-PC with the lipid bilayer induces a conformational change. Rupture of these MscL containing vesicles at a PEG-DSPE coated solid-substrate, creates a flexible MscL containing bilayer which we structurally characterise with polarised neutron reflectometry (PNR). By using a new tethered lipid bilayer system we were able to incorporate unclustered MscL at a high volume fraction. The PNR from this bilayer before and after exposure to the antimicrobials is consistent with an upward movement of the C-terminal protrusion, partially into the proteins transmembrane domain on MscL gating that has been predicted by Molecular Dynamics, based on electron paramagnetic resonance (EPR) and fluorescence resonance energy transfer (FRET) measurements. \cite{deplazes2012structural}

%%%%%%%%%%%%%%%%%%%%%%%%%%%%
\section{Materials and Methods}
%%%%%%%%%%%%%%%%%%%%%%%%%%%%

%%%%%%%%%%%%%%%%%%%%%%%%%%%%
\subsection{Materials}
%%%%%%%%%%%%%%%%%%%%%%%%%%%%
%\begin{sloppypar}
POPC (1-palmitoyl-2-oleoyl-glycero-3-phosphocholine, 850457P, Avanti Polar Lipids (Alabaster, AL, USA))   and 
POPG (1-palmitoyl-2-oleoyl-sn-glycero-3-phospho-(1-rac-glycerol) (sodium salt), 840457P, Avanti Polar Lipids (Alabaster, AL, USA)  )  lipids and 
DSPE\-PEG2000\-PDP tethers (1,2-distearoyl-sn-glycero\-3-phosphoethanolamine\-N-[PDP(polyethylene glycol)-2000] (ammonium salt), 880127P, Avanti Polar Lipids (Alabaster, AL, USA)  ) were purchased as solid powders. Components for buffer solution and chloroform, ethanol were purchased from Sigma Aldrich (Dorset, UK). The peptide Pexiganan was purchased from China Peptides (Shanghai, China) with a purity of greater than 98$\%$ .

%\end{sloppypar}

\subsection{Vesicle preparation pre-protein expression}
3:1 POPC:POPG MscL containing vesicles were prepared by dissolving a 3:1 mixture of POPC and POPG lipids in the minimum amount of chloroform, evaporating the chloroform under a steady stream of nitrogen and rehydrating the lipid film in HEPES buffer (20~mM, pH~7.4, 150~mM NaCl). To ensure unilamellar vesicles (liposomes) of diameter $\sim$100~nm, the liposome solution was tip-sonicated using a Sonics Vibra Cell model VCX 500 with a CV33 converter from Sonics and Materials Inc. (Newtown, CT, USA) (5 seconds on, 5 seconds off at 20\% power) for 30 minutes until translucent.

%%%%%%%%%%%%%%%%%%%%%%%%%%%%
\subsection{Cell-free expression of MscL}
%%%%%%%%%%%%%%%%%%%%%%%%%%%%
The MscL expression plasmid, a pDuet-1 WT MscL-6 His construct under
T7 promoter, was kindly supplied by Paul Rohde and Boris Martinac of
the Victor Chang Cardiac Research Institute, Sydney. The cell free protein
expression was carried out using RTS500 Biotechrabbit Proteomaster E. coli
HY kit (Biotechrabbit GmbH, Berlin,Germany) following manufacturer instructions.\cite{biotechrabbit2023} Optimisation of the expression was done with further guidance taken from Abdine et al., on their optimisation
of MscL expression.\cite{abdine2011cell} The expression reaction solution contained 4~mg of 3:1
POPC:POPG lipid per 1 mL reaction mix, as the membrane construct the protein
is expressed into. The embedded protein composition of the resulting vesicles was determined to be 15\% (w/w) by solubilising the protein in Triton X-100 (Sigma Aldrich, Dorset, UK) for removal of the lipids and then quantification using the Bicinchoninic acid (BCA) method. Further information can be found in the Supplementary Information (SI Section 4).

%%%%%%%%%%%%%%%%%%%%%%%%%%%%
\subsection{Substrate preparation}
%%%%%%%%%%%%%%%%%%%%%%%%%%%%
%\begin{sloppypar}
\textbf{Tether solution.} The tether solution  contained 0.1~mg/mL of 1,2-distearoylsn-glycero-3-phosphoethanolamine-N-PDP (polyethylene glycol)2000] (ammonium salt) (DSPE-PEG2000-PDP) in ethanol.

\noindent
\textbf{QCM-D substrates.} For the QCM-D measurements, gold coated quartz-crystal sensors (QSX-301 Gold) were purchased from QSense (Biolin Scientific/Q-Sense AB, Västra Frölunda, Sweden). These sensors were cleaned by the RCA method (submerging in a 1:1:50 mixture of NH$_4$OH:H$_2$O$_2$:H$_2$O, at 55~$^\circ$C for 10 minutes), rinse with copious Milli-Q water, followed by drying with nitrogen and UV/ozone cleaning for 10 minutes.  Sensors were submerged in tether solution in a glass beaker, in the dark at 4~$^{\circ}$C for 12 hours. The sensors were rinsed with Milli-Q water before loading into the flow cells and the underside of the sensor dried with nitrogen before the cell was assembled.

\noindent
\textbf{NR substrates.} For the NR measurements, ozone cleaned silicon crystals ($50 \times 80 \times 15$ mm) with a polished $80 \times 50$ mm face (111 orientation, surface roughness (RMS) $<$8~\AA{}) were purchased from PI-KEM. These silicon blocks were cleaned in piranha solution (3:1:5 sulphuric acid, 30\% hydrogen peroxide and water), rinsed in a copious amount of Milli-Q water, dried under a stream of nitrogen and ozone-cleaned. The silicon blocks were determined to have less than 8~\AA{} roughness prior to sending to the NIST nanofabrication
facility, where they were sputter-coated with permalloy (Ni80Fe20) and gold (approximately 15~nm thickness each) at the NIST center for Nanoscience and Technology, Gaithersburg, MD, U.S.A., in a Denton Discovery 550 sputtering chamber. Prior to the NR measurments, the substrates were exposed to UV/ozone for \textgreater10~min,  rinsed immediately with Milli-Q water, then dried in a gentle nitrogen stream and immediately submerged in tether solution for 12~h. They were then gently rinsed with ethanol and Milli-Q water before the measurement.
%\end{sloppypar}
%%%%%%%%%%%%%%%%%%%%%%%%%%%%
 \subsection{Quartz Crystal Microbalance with Dissipation (QCM-D)}\label{sec:QCM}
%%%%%%%%%%%%%%%%%%%%%%%%%%%%
QCM-D measurements were performed on the Biolin Scientific E4 Q-Sense Instrument at
ISIS, Oxford, UK. QCM-D sensors were cleaned as described above. The QCM-D cells were cleaned prior to use as per manufacturer instruction and all tubing used was new and rinsed through with 20~mL of ethanol and 20~mL of Milli-Q water prior to use.
Buffer and liposome solutions were introduced to the sensor using a peristaltic pump
flowing in the solution at 0.1~mL/min. Figure~\ref{fgr:QCM-D} plots the shift in the frequency of the third overtone, $\Delta F$, and the dissipation, $\Delta D$, which are determined by the instrument software, during bilayer formation. The latter is the reciprocal of the oscillator's Q factor and provides a measure of the viscoelasticity of the layer. Further details can be found in the Supplementary Information (SI Section 1).

%%%%%%%%%%%%%%%%%%%%%%%%%%%%
\subsection{Neutron reflectivity}
%%%%%%%%%%%%%%%%%%%%%%%%%%%%
Polarised Neutron reflectometry (PNR) measurements were carried out using the PolRef time-of-flight reflectometer at the ISIS Neutron and Muon source (experiments RB1920647 and RB1820534).\cite{POLREF_rb1920647, POLREF_rb1820534, Ayscough_Tethered_Lipid_Bilayers_2025} A broad band neutron beam with wavelengths from 2 to 12~\AA{} was used.
The reflected intensity is measured as a function of the scattering vector $Q_z=\frac{4\pi}{\lambda}\sin(\theta)$, where $\lambda$ is wavelength and $\theta$ is the incident angle. The collimated neutron beam was reflected from the solid-liquid interface at different glancing angles of $\theta=0.5, 1.2$ and $2.5^{\circ}$ in order to cover the desired $Q$ range, i.e. from total reflection edge to background. The permalloy layer in our samples was magnetized parallel to the neutron polarisation vector to saturation in a static magnetic field. This modifies the refractive indices for the neutron depending on the neutron polarisation state (up or down) - producing different scattering cross-sections (SLDs) for the two spin states, resulting in different reflectivities for spin up and spin down.
This method has previously been referred to as magnetic contrast. The advantages of this approach have been discussed previously.\cite{Treece2019, hughes2019physical, LeBrun2008}
The presence of a (high SLD) permalloy layer in our system increases the reflectivity (at low $Q_z$) which increases the signal-to-background ratio. This effect is
independent of any magnetic contrast. In addition, we obtain two separate measurements for each sub-phase contrast, which provides further constraints to the fitting parameter space.

Purpose-built liquid flow cells for analysis of the silicon-liquid interface were placed on
a  sample stage mounted on a goniometer in the NR instrument and the inlet to the liquid cell was connected to a liquid chromatography pump (JASCO PU-4180), which allowed the automated
exchange of the solution isotopic contrast  within the (3~mL volume) solid-liquid sample cell.
For each solution isotopic contrast change, a total of 20~mL solution (D$_2$O(l), H$_2$O(l) or gold matched buffer (GMW(l)) was pumped through the cell at a speed of 1.5~mL/min. The sample temperature was held at 20~$^{\circ}$C using a recirculated water bath.

%%%%%%%%%%%%%%%%%%%%%%%%%%%%
\subsection{Neutron reflectivity data analysis}
%%%%%%%%%%%%%%%%%%%%%%%%%%%%
Neutron reflectivity data were analyzed using the RefNX software package\cite{Nelson2019}, which employs an optical matrix formalism\cite{Born1997} to fit layer models representing the interfacial out-of-plane structure. In this approach the interface is described as a series of slabs, each of which is characterized by its scattering length density (SLD), thickness, roughness and hydration if applicable. For each layer,$L$, in the model, the scattering length density $\rho_L=\sum_i\phi_i\rho_i$, where $\phi_i$ and $\rho_i$ are the volume fraction and scattering length density of component $i$, with the sum over all the components present in that layer. Interfacial roughness was implemented in terms of an error function, according to the approach by Nevot and Croce\cite{NevotL.1980}. 

The final model  comprised  the following distinct layers on a silicon substrate:  silicon oxide, permalloy, gold, PDP, PEG, DSPE, MscL containing lipid bilayer and MscL protrusion. In the model for the bilayer, the scattering length density and thickness of each layer are coupled through the area per lipid molecule (APM), which is constrained to be same in the head/tail regions, ensuring each lipid head group is coupled to a diacyl tail, but is allowed to differ between  the inner and outer leaflets. The bilayer coverage gives the fraction of the neutron beam footprint area that is covered by protein-containing bilayer, with the complement being water, and the protein coverage gives the fraction of that bilayer area that is occupied by protein. Compared to the tail group layers we allow the head group layers to incorporate additional water molecules per lipid head (WPLH). Pexiganan is not explicitly included in the model.  Our previous studies indicated a low peptide to lipid ratio,\cite{mckinley2017} meaning that the small contribution to the scattering length density of the layer does not warrant the concomitant increase in the complexity of the model. A full description of the model used to fit the data is provided in the Supplementary Information (SI section 2) with its implementation in Python (SI section 6).

A differential evolution minimization was used to adjust the fit parameters to reduce the differences between the model reflectivity and the data. In all cases the simplest possible model (i.e. fewest layers), which adequately described the data, was selected. Error analysis of the fitted parameters was carried out using RefNX's “Bayesian” error algorithm. For the sampling, we used 800 “burn in” points followed by 4000 samples with a thinning of 100. The resulting plots contain fits and corresponding real space structure of the sample layer system, as well as 300 samples from the posterior distributions (shown as shaded lines/regions).

%%%%%%%%%%%%%%%%%%%%%%%%%%%%%%%%%%%%%%%%%%%%%%
\subsection{Small Angle Neutron Scattering}
%%%%%%%%%%%%%%%%%%%%%%%%%%%%%%%%%%%%%%%%%%%%%%%

Small angle neutron scattering (SANS) experiments were carried out on SANS2D, ISIS Neutron and Muon source (experiment RB180511).\cite{SANS2D_rb1820511} All samples were measured with an \num{8} \si{\milli\m} aperture with a source-to-sample and sample-to-detector distance of \num{12} metres whilst being held at \num{20} \si{\degreeCelsius} using a recirculated water bath. 
For samples suspended in D$_2$O buffer, exposures of \num{12} mAh ($\sim$ 21 minutes exposure) and \num{8} mAh ($\sim$ 14 minutes exposure) proton beam current were used, for SANS and transmission measurements respectively. Rectangular cross--section quartz cuvettes of \num{1} \si{\milli\metre} path length were used for H$_2$O contrast and \num{2} \si{\milli\metre} path length cuvettes were used for D$_2$O contrast buffer.

%%%%%%%%%%%%%%%%%%%%%%%%%%%%%%%%%%%%%%%%%%%%%%%%%%%%
\subsection{SANS fitting}
%%%%%%%%%%%%%%%%%%%%%%%%%%%%%%%%%%%%%%%%%%%%%%%%%%%
The small angle scattering from a system of particles can be characterised by Guinier (low-$Q$) and Porod regions (high-$Q$). The Guinier region allows a radius of gyration of the associated scatterer to be determined whereas the Porod region provides information on the shape of particles and the roughness of the scattering surface.

The SANS data were fitted to two level Guinier-Porod models (see SI section 5) using the Irena and Nika SAS fitting packages in Igor Pro.\cite{GPmodelirena}

%%%%%%%%%%%%%%%%%%%%%%%%%%%%
\section{Results and discussion}
%%%%%%%%%%%%%%%%%%%%%%%%%%%%

\subsection{Characterisation of the MscL containing vesicles}

%%%%%%%%%%%%%%%%%%%%%%%%%%%%

MscL containing vesicles were produced using cell free protein expression (details in Experimental section) in  which the protein is assembled directly into vesicles of a chosen lipid composition of 3:1 POPC:POPG.

\begin{figure}
\centering
\includegraphics[width=\linewidth, scale=0.2]{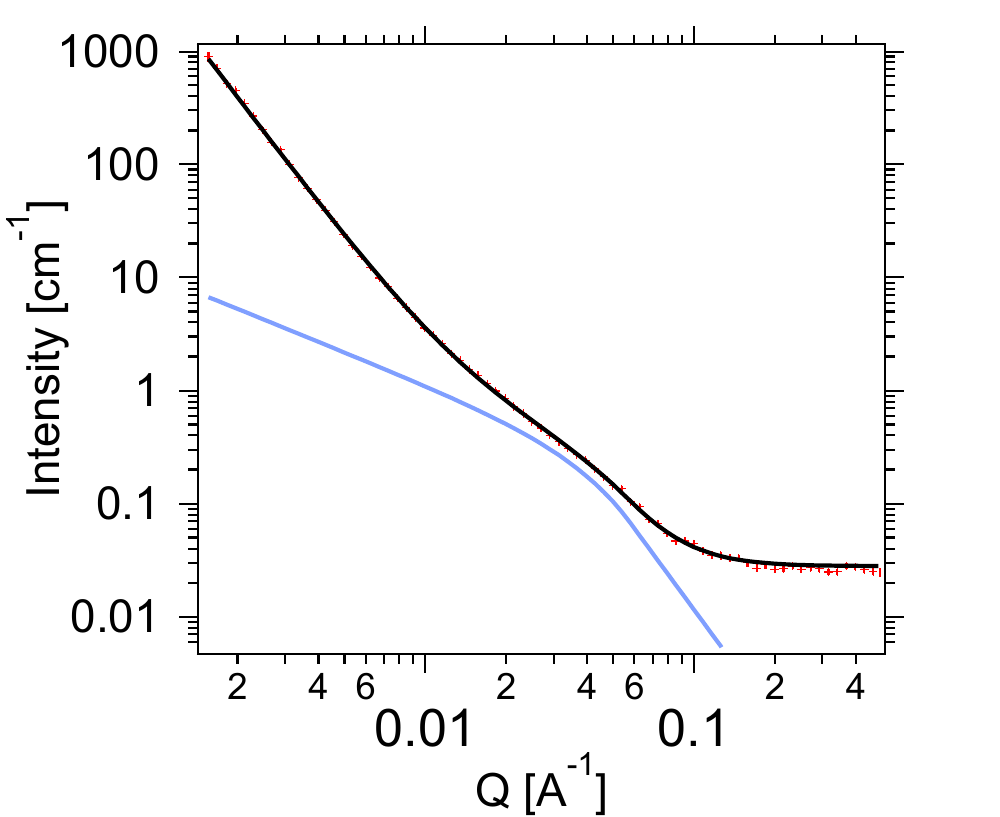}
  
  \caption{Small angle neutron scattering (SANS) 
  of MscL containing vesicles measured in D$_2$O 20 mM 
  HEPES buffer pD 7.4. 2-level Guinier-Porod model shown 
  in as solid black line, experimental SANS data is displayed 
  as red error bars and the blue curve displays the contribution 
  of the first Guinier-Porod level to the SANS curve, 
  which we attribute to  the scattering from the individual 
  MscL proteins. }
  \label{d2omolfr}
\end{figure}

The vesicles were characterised using small angle neutron scattering. In D$_2$O buffer there is significant contrast between the buffer and both the lipid and protein components of the vesicles meaning that the scattering is sensitive to both the lipid bilayer and the proteins. The small angle scattering data is shown in Figure \ref{d2omolfr} and the fitted parameters are in Table \ref{GPtable}. A two-level Guinier-Porod (GP) model was used to fit the scattering curves, accounting for the scattering from the overall shape and size of the vesicle (scattering contribution to the low $Q$ region) and from the embedded proteins (high $Q$ region).

The Porod exponent of $d=3.14\pm 0.08$ for the second GP level (GP2) suggests that the interface between the protein containing vesicle and the buffer is rougher than  POPC vesicles, for which a Porod exponent of $d$=3.6 has previously been reported.\cite{nele2019effect} This increased roughness is consistent with that observed by cryo-Electron Microscopy on proteoliposomes prepared using a similar cell-free expression.\cite{Berrier2011} As MscL is a double-spanning  transmembrane protein and in detergent-free cell-free expression the protein is synthesized outside the vesicle, the  C-terminus region of MscL is expected to protrude out from the lipid bilayer.

The radius of gyration (GP1) of the protein was determined to be 26$\pm$3~\AA{}, in agreement with the theoretical value of 28~\AA{}, which was estimated from the crystal structure 2OAR\cite{Rees2007}. This strongly suggests that this contribution to the scattering curve is from individual MscL channels within the vesicle bilayer. This contrasts with the large clusters of MscL observed using SANS when the MscL was produced by bacterial over-expression and reconstituted into DOPC vesicles.\cite{GrageBPJ11}. As the protein will only assemble into an ion channel in the presence of lipids/surfactants and our cell-free expression is surfactant-free, we can be sure that the MscL is embedded in the vesicles. Although the direct measurement of ion channel activity and gating is beyond the scope of this study, patch clamp recording has previously been used to demonstrate that channels inserted into liposomes using a similar cell-free expression protocol are functional.\cite{Berrier2011} In the Supplementary Information (SI section 5.2) we show that the best fit radius of gyration for the protein increases in response to the addition of lyso-PC, which has previously been shown to gate MscL open,\cite{martinac1990mechanosensitive,perozo2002open,Strutt21} demonstrating that the protein has been expressed into the vesicles in an active form, capable of undergoing a conformational change triggered by interaction with amphipathic molecules. We further observe a change in scattering on addition of pexiganan, shown in the S.I., also suggestive of a conformational change but inconclusive due to a change in scattering of the overall vesicle and protein. We demonstrate in SI section 5.1, that the observed changes in radius of gyration could be consistent with a change from a closed to an open channel conformation.

\begin{table*}
\centering
\caption{Best fit parameters of 2-level Guinier-Porod model for SANS measured from MscL containing vesicles in D$_2$O buffer shown in Figure \ref{d2omolfr}. }
\begin{tabular}{|c|l|l|}
\hline
Parameter & GP1 &	GP2	\\
\hline
$s$	& 0.96 $\pm$ 0.3 &	0 (fixed)\\
$G$	& $(1.33 \pm$ 0.17)$\times10^{-2}$ &	(3.9$ \pm 0.1)\times 10^{-12}$ \\
$R_g$ (Å)	& 26.1 $\pm$ 2.6 &	10$^6$ (fixed) \\
$d$	& 3.2 $\pm$ 0.3 & 3.14 $\pm$ 0.08 \\
$\chi^2$	& 72.1 &  \\
Normalised $\chi^2$	& 1.08 &  \\
\hline
\end{tabular}
\label{GPtable}
\end{table*}

%%%%%%%%%%%%%%%%%%%%%%%%%%%%
\subsection{Characterisation of the tether layer}
%%%%%%%%%%%%%%%%%%%%%%%%%%%%

\begin{figure*}[ht]
\includegraphics[width=\textwidth]{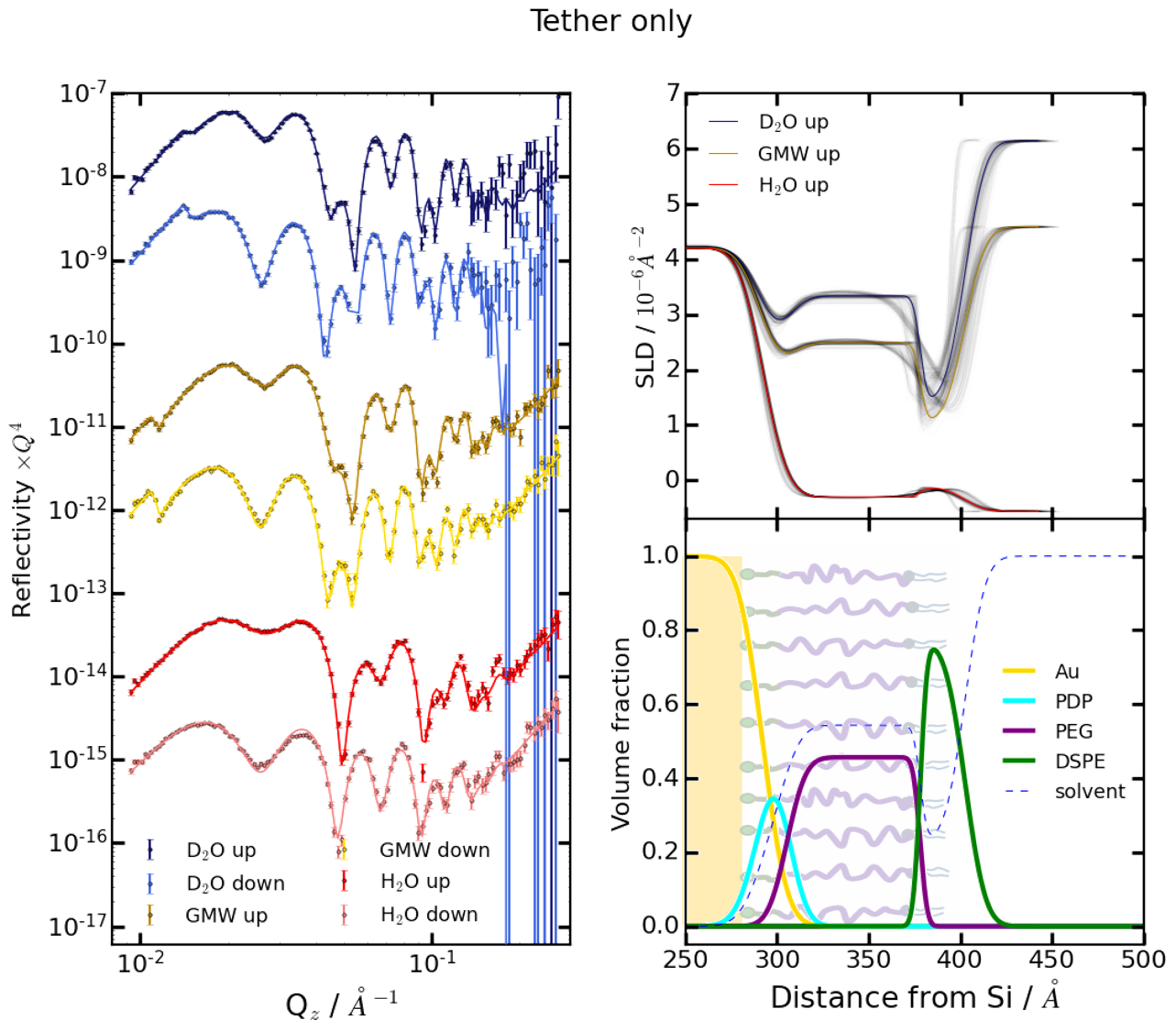}
  \caption{Fitted neutron reflectivity of PDP-PEG2000-DSPE tethered gold surface, measured in three solvent contrasts (D$_2$O, H$_2$O and gold matched water) and two spin contrasts (left panel). Corresponding SLD profiles (top right panel). Calculated volume fraction profiles (bottom right panel). The shaded areas represent 300 samples from the MCMC sampling, as given by RefNX. The faded cartoon in the bottom right panel depicts the arrangement of the gold, PDP, PEG, DSPE layers, and the solvent, from left to right.}
  \label{fgr:TetherOnly}
\end{figure*}

Functionalised NR substrates were measured using PNR in three buffer contrasts: D$_2$O(l), a mixture of D$_2$O/H$_2$O contrast matched to the SLD of gold (GMW(l)) and  H$_2$O(l). Unless otherwise stated, the buffer used is HEPES buffer (20~mM, pH/pD 7.4, 150~mM NaCl). The SLD of the permalloy layer between the gold and silicon differs for spin up and spin down neutrons, providing two magnetic contrasts  for each of the three buffer contrasts. Data are displayed in the left panel of Figure~\ref{fgr:TetherOnly} as reflectivity$\times$Q$^4$ vs. $Q$ to highlight differences measured between the contrasts and differences between the tether and bilayer layers, rather than $Q^{-4}$ decrease in the reflectivity caused by the   presence of the solid interfaces. For clarity, only the SLD profiles for spin up neutrons are shown in the upper right of Figure~\ref{fgr:TetherOnly}. The volume fraction profiles for each component are shown in the lower right panel of Figure~\ref{fgr:TetherOnly}.

The tether was modelled by three layers: one for the PDP linker to the gold surface, one for the PEG spacer and one for the DSPE lipid. The small component volume of the ethanolamine head group and the interfacial roughness of the brush-like tether mean that it is not appropriate to treat the lipid head group and tails as separate layers. .

From the MCMC analysis of our tether model, it was determined that the thickness of the solid layers could be determined within one \AA{} (SI, Table~S1). The uncertainty on the parameters of the tether layers is higher, which is partly due to a high level of hydration and low contrast. 

In contrast to a typical polymer brush, in which the interfacial width is a consequence of the the distribution of chain ends throughout the brush \cite{Milner1988}, the capping of the chain ends by lipid moieties provides a driving force for these lipid tail end caps to segregate to the top of the brush layer. This results in a lower interfacial roughness for the tether/water interface than is typical for a brush/solvent interface. 
The best fit to the solid parameters derived from the fitting of our model to the the reflectivity data, were used to constrain the fits for the subsequent bilayer analysis.

%%%%%%%%%%%%%%%%%%%%%%%%%%%%
\subsection{Formation and characterisation of a tethered MscL/lipid bilayer}
\begin{figure*}[ht]
\centering
\includegraphics[width=0.5\linewidth]{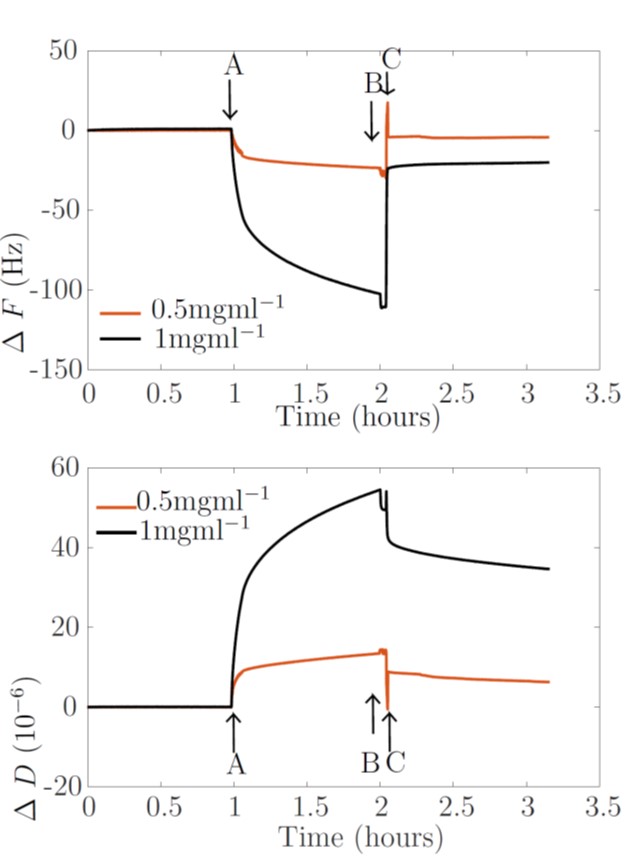}
  \caption{QCM-D measurement of the changes in (3rd Overtone) Frequency (upper
Figure) and  Dissipation (lower Figure) of a DSPE-
PEG2000-PDP tethered gold sensor after addition of vesicles (A)
at 0.5~mg/mL (red line) and 1~mg/mL (black line),
300~mM NaCl injected at point B followed by 150~mM NaCl at point C.}
  \label{fgr:QCM-D}
\end{figure*}

%%%%%%%%%%%%%%%%%%%%%%%%%%%%
The bilayer formation was optimised prior to the NR measurements using QCM-D and is dependent on vesicle rupture, which is driven by changes in the balance between vesicle-substrate adhesion, the bending energy of the lipid bilayer and the osmotic stress across the bilayer.  A 1~mg/mL  solution of MscL containing vesicles produced a frequency shift of ca. 100~Hz and was accompanied by an increase in dissipation of ca. 50$\times 10^{-6}$, corresponding to the adhesion of intact protein-containing vesicles, likely driven by the insertion of DSPE at the end of the PEG tether into the outer leaflet of the vesicle. Subsequent injection of NaCl solutions (300~mM NaCl, followed by 150~mM NaCl) provide the osmotic stress that ruptures the vesicles,  resulting in a net frequency shift of ca. 10~Hz (Figure~\ref{fgr:QCM-D}), which confirmed
the formation of a dense bilayer. In the literature, $-\Delta D/\Delta F$ ratios have been used to reveal the nature of the vesicle-substrate interaction and specifically identify  bilayer formation. For the 1~mg/mL solution we observed $-\Delta D/\Delta F\approx 5$, which lies in the range 3 to 6, which has been reported for complete bilayer formation in tethered systems~\cite{INCI2015}.
The large decrease in frequency and increase in dissipation that occurs on the addition of MscL containing  vesicles  is less pronounced at a lower vesicle concentration (indeed the frequency shift for the 0.5~mg/mL case returns to zero). The variation in $(\Delta F,\Delta D)$ observed in Figure~\ref{fgr:QCM-D} can be explained by: adsorption of intact MscL containing  vesicles from A$\rightarrow$B;  osmotically induced rupture of the MscL containing  vesicles to form a bilayer in the case of a high coverage of adsorbed MscL containing  vesicles (at 1 mg/mL) and desorption of the adsorbed MscL containing  vesicles when they are below a critical coverage (at 0.5 mg/mL). 

\begin{figure*}[ht]
\includegraphics[width=\textwidth]{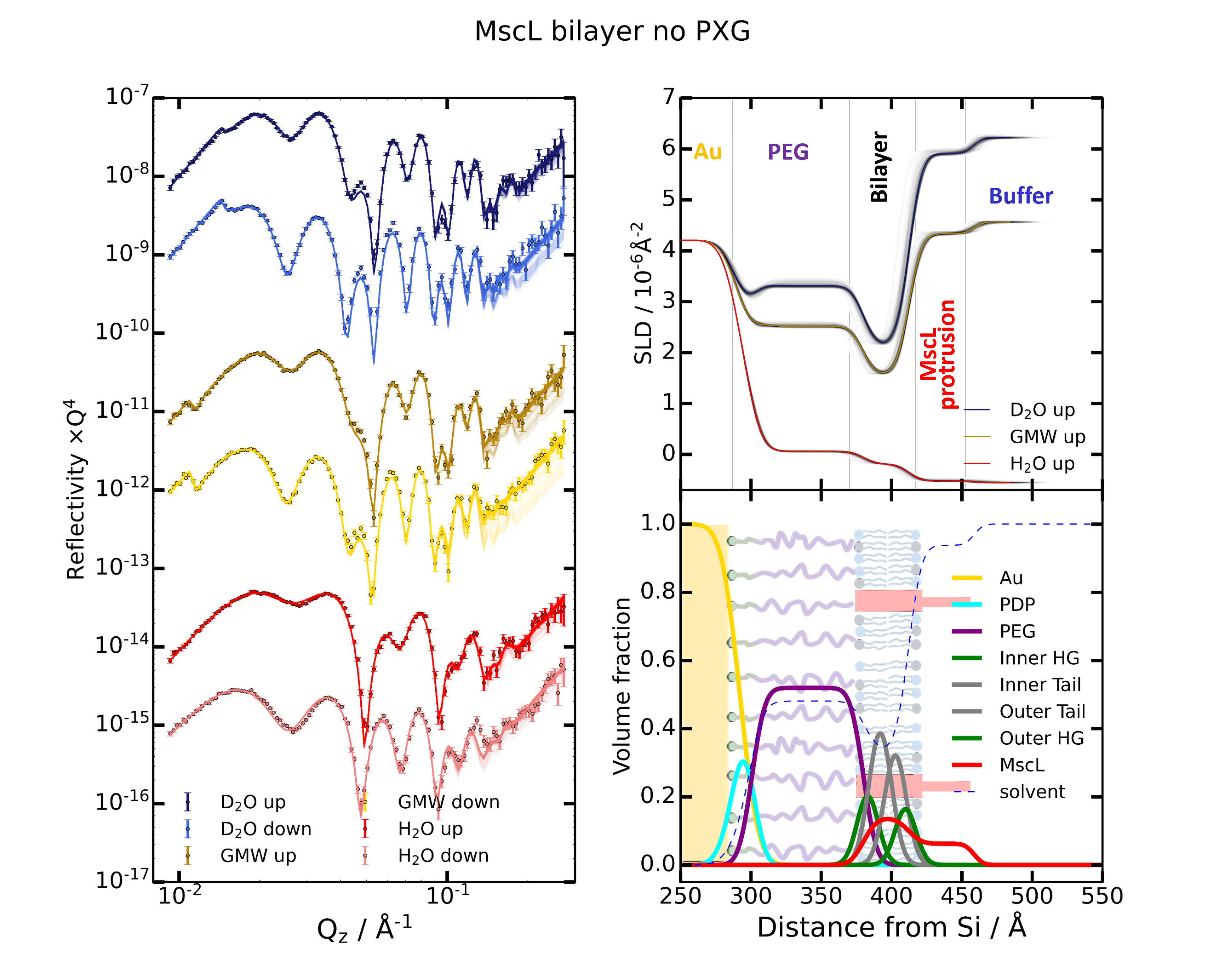}
  \caption{Fitted neutron reflectivity of PDP-PEG2000-DSPE tethered bilayer formed from 3:1 POPC:POPG (MscL containing) vesicles measured in three solvent contrasts (D$_2$O, H$_2$O and gold matched water) and two spin contrasts (left panel). Corresponding SLD profiles (top right panel). Calculated volume fraction profiles (bottom right panel). The shaded areas represent 300 samples from the MCMC sampling, as given by RefNX.}
  \label{fgr:NR_MscL_bilayer}
\end{figure*}

For the structural measurements (NR), MscL containing vesicles (18~mL at 1~mg/mL) were injected into the neutron reflectivity cells
using a syringe pump and incubated for 1~hour. Vesicle rupture by osmotic shock was achieved by injecting 20~mL of 300~mM NaCl containing buffer followed by 20~mL of 150~mM NaCl containing buffer. The resulting membrane mimetic 
was then measured in three buffer contrasts (D$_2$O, GMW, H$_2$O), see Figure~\ref{fgr:NR_MscL_bilayer}. 

%\begin{sloppypar}
Bilayer formation by rupture of the MscL containing vesicles incorporates the DSPE part of the tether into the inner leaflet of the bilayer and so the tethered bilayer model does not have a distinct DSPE layer. 
For the two lipid leaflets of the lipid bilayer, separate area per molecule (APM) and water per lipid heads (WPLH) parameters were fitted. The best fit APM values for the inner and outer lipid bilayer leaflets were 79$\pm$2~\AA{}$^2$ and 98$\pm$3~\AA{}$^2$, respectively, both with 6$\pm$1 WPLH. 
%\end{sloppypar}

The tether leaflet APM of 79~\AA$^2$ agrees well with the literature measurement for a similar mixed DSPE-PEG(2000)/DPPC  monolayer at $\Pi$=30 mN/m, which is often taken as the canonical surface pressure associated with lipid bilayer leaflets.\cite{Kowalska21}
Our model also included an additional layer to account for the protrusion of the C-terminus containing domain of the MscL ion channel observed in other studies~\cite{chang1998structure}. To avoid bias, we tested models with protrusions on either and both sides of the bilayer. The model that provides the best fit to the data was one with the protrusion facing away from the substrate (see  Figs.~\ref{fgr:NR_MscL_bilayer} - \ref{fgr:NR_MscL_bilayer_3p2_PXG}).

The best fit bilayer coverage indicates that there are areas of the PDP-PEG brush that are not covered by bilayer. As the open channels occlude a larger area than the closed channels, it could be that the presence of such defects within the bilayer is necessary to accommodate the concomitant areal expansion of the bilayer with embedded channels without inducing a buckling.
The coverage of MscL within the lipid bilayer is 14$\pm$1\%, which is within error of that expected based on the composition of the protein-containing vesicles that rupture to form the bilayer.  
The thickness of the MscL protrusion from the lipid bilayer (Table~\ref{tbl:bil_PXG}) was fitted to be 46$\pm$3~\AA{} and the roughness of the final bilayer is $\sim$8~\AA{}, which is consistent with a bilayer formed on-top of a flexible polymer layer.

Given the PEG thickness, $t$=80$\pm $1~\AA{}, volume fraction, 
$\phi_{\mathrm{PEG}}=0.44$ and that the molecular volume of PEG2000 is $v_{\mathrm{PEG}}=2966.7$~\AA{}$^3$,
the APM of the PEG tether can be calculated to be 84$\pm$1~\AA{}$^2$ 
($APM_{\mathrm{PEG}}=v_{\mathrm{PEG}}/(\phi_{\mathrm{PEG}}t$). From the APM of a PEG chain, a distance between the tether sites can be estimated as
$D = 2\sqrt{\frac{APM}{\pi}}=10.4$ \AA{}. As the distance between tether sites is significantly lower than the Flory radius of 33~\AA{}, the PEG tether will be in the brush regime.

As can be seen from the difference between the APM for the inner and outer leaflets (Table~\ref{tab:PXG_bil}), the tether density has imposed some asymmetry in the leaflet packing densities. The multi-component character of the lipid bilayer means that there may be some compositional asymmetry, which in principle could be investigated using selective deuteration. We did not pursue this as our focus was to determine the protein composition of the bilayer and to observe conformational changes induced by antimicrobials. Some of the consequences of structural asymmetry on membrane elastic and thermodynamic properties have recently been reviewed.\cite{Deserno23} We also note that adopting a conical model for the p7 porin, incorporated in a POPC bilayer resulted in an asymmetry in the head group layer thicknesses of the inner and outer leaflets.\cite{Watkins17} 
The outer leaflet APM of 99~\AA$^2$ is larger than than the value of 65~\AA$^2$ found by Molecular Dynamics simulation for a 7:3 POPC/POPG bilayer,\cite{Agrawal23} but lies between the values determined for an MscL incorporating bilayer without tension and under tension.\cite{Melo17} To date there is little experimental information on the APM for lipids in bilayers that incorporate transmembrane proteins. At 15\% (w/w) MscL each protein is associated with about 260 lipids/leaflet, and Molecular  Dynamics simulations have shown the C-terminus side of the protein to be hydrogen-bonded to 40 lipids,\cite{Arroyo14} which could result in a thinner, more disordered, lipid layer than in the absence of a protein.\cite{Jeon08}

In conclusion, we have demonstrated that a lipid bilayer, containing the bacterial membrane protein MscL, tethered to a gold substrate by PDP-PEG2000-DSPE molecules, can be formed by deposition and rupture of MscL containing POPC:POPG (MscL containing) vesicles. Our measurements clearly show the successful incorporation of the transmembrane protein MscL, with a final protein volume coverage of 14$\pm$1\%. 

%%%%%%%%%%%%%%%%%%%%%%%%%%%%
\subsection{MscL/lipid bilayer interaction with pexiganan (PXG)}
%%%%%%%%%%%%%%%%%%%%%%%%%%%%
Minimum inhibitory concentrations (MICs) of the AMP pexiganan for aerobic and anaerobic bacterial strains  have been reported to be in range of 0.4 to 12.8~$\mu$M~\cite{Gottler2009}. 
The response of the MscL containing bilayer to the AMP pexiganan was tested for two
concentrations in the lower part of this range, by sequentially flowing 1.6~$\mu$M PXG in D$_2$O HEPES buffer and 3.2~$\mu$M PXG in the same buffer through the NR cell. Pure buffer rinses (20~mL) were used between and after these steps to ensure that cumulative deposition was not occurring during the NR experiments.

\begin{figure*}[ht]
\includegraphics[width=\textwidth]{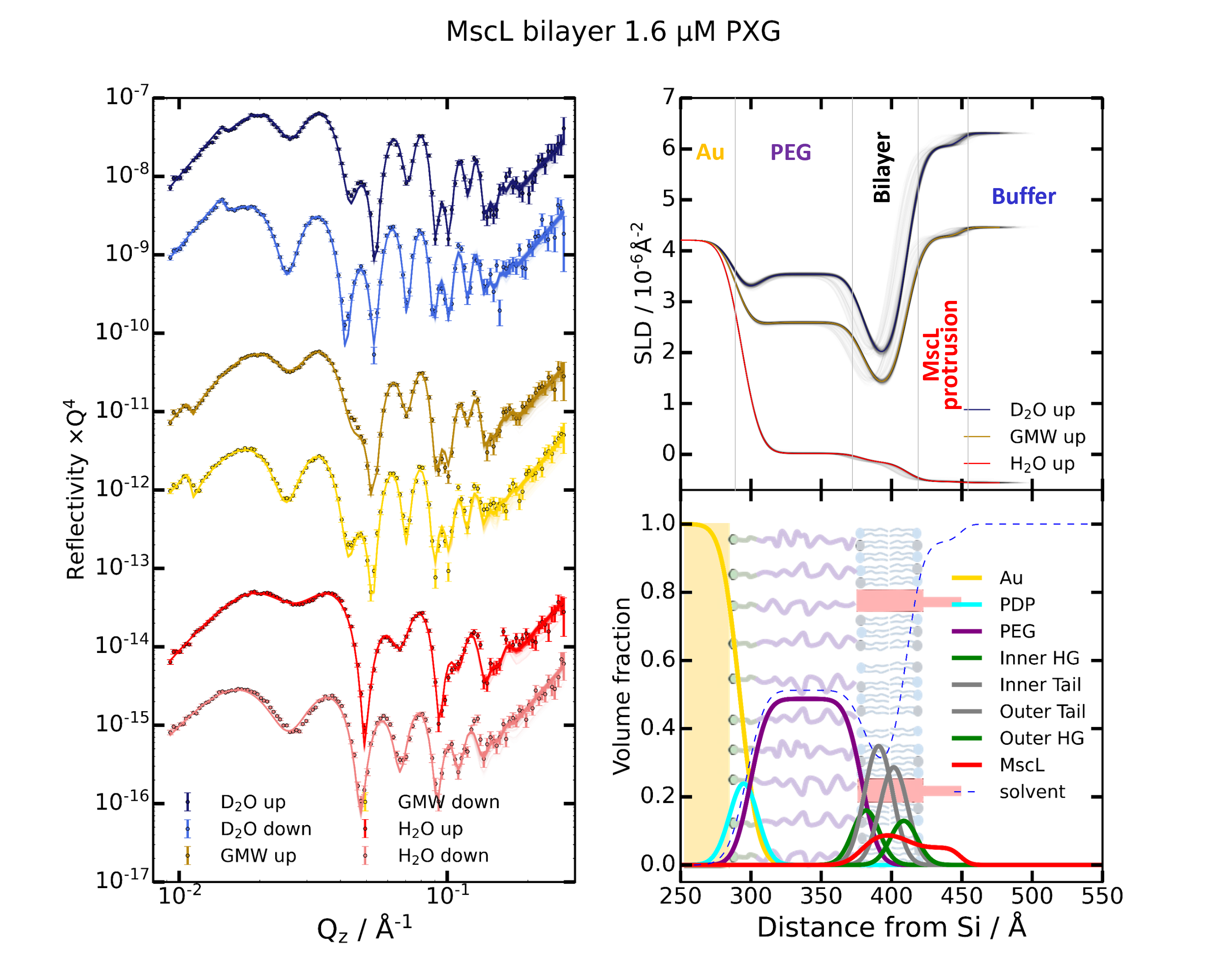}
  \caption{Fitted Reflectivity profiles of tethered bilayers containing MscL after the addition of pexiganan at 1.6~$\mu$M. The six curves were measured in three solvent contrasts (D$_2$O, H$_2$O and gold matched water) and two spin contrasts (left panel). Corresponding SLD profiles (top right panel). Calculated volume fraction profiles (bottom right panel). The shaded areas represent 300 samples from the MCMC sampling, as given by RefNX. }
  \label{fgr:NR_MscL_bilayer_1p6_PXG}
\end{figure*}

Changes in the reflectivity data could be observed after PXG addition: by comparing Figure~\ref{fgr:NR_MscL_bilayer} with Figure~\ref{fgr:NR_MscL_bilayer_1p6_PXG}, differences can be seen between the reflectivities of the tethered bilayer before and after PXG addition, resulting in changes to the SLD profile in the MscL protrusion region. These differences are most apparent in the GMW and H$_2$O contrasts (e.g. fringes at $Q$ values of 0.1~\AA{}$^{-1}$ and beyond become more pronounced). The hydration of the PEG layer (purple line) slightly increases (from ca. 48 to ca. 51\%) and the volume fraction distribution of the MscL (red line) changes.

\begin{figure*}
\includegraphics[width=\textwidth]{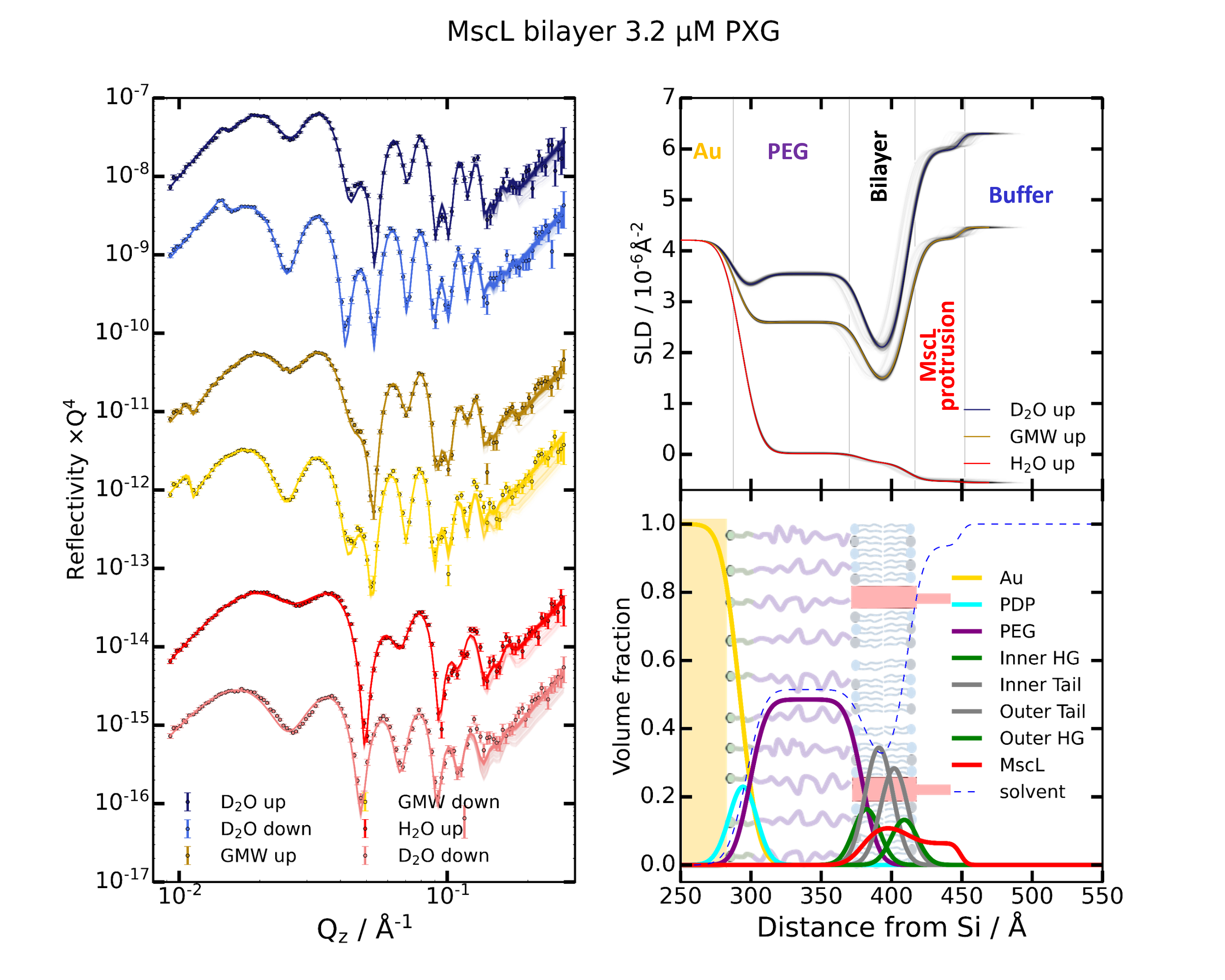}
  \caption{Fitted Reflectivity profiles of tethered bilayers containing MscL after the addition of pexiganan at 3.2~$\mu$M. The six curves were measured in three solvent contrasts (D$_2$O, H$_2$O and gold matched water) and two spin contrasts (left panel). Corresponding SLD profiles (top right panel). Calculated volume fraction profiles (bottom right panel). The shaded areas represent 300 samples from the MCMC sampling, as given by RefNX. }
  \label{fgr:NR_MscL_bilayer_3p2_PXG}
\end{figure*}

The difference between the initial bilayer and 1.6~$\mu$M PXG is larger than seen with subsequent addition of 3.2~$\mu$M PXG (Figure~\ref{fgr:NR_MscL_bilayer_3p2_PXG}), though some differences are observed with the increased PXG concentration at high $Q$. A comparison of the best fit parameters for the bilayer and MscL protrusion are shown in Table~\ref{tbl:bil_PXG}. The parameters corresponding to solid layers were constrained to the values corresponding to the best fit to the tether data (Table~S1). 

As described in the Experimental section, we have conducted a detailed error analysis in order to quantify the significance of the observed parameter changes: in addition to the fitted parameter values, we have also calculated their Bayesian posterior distribution, which expresses the probability of a parameter taking a value given the experimental evidence. In Bayesian statistics, the posterior probability is proportional to the product of likelihood and prior probability. Thus, a reduction in the width of the posterior compared to the prior distribution implies a significant gain of information.

\begin{figure}
\includegraphics[width=\linewidth]{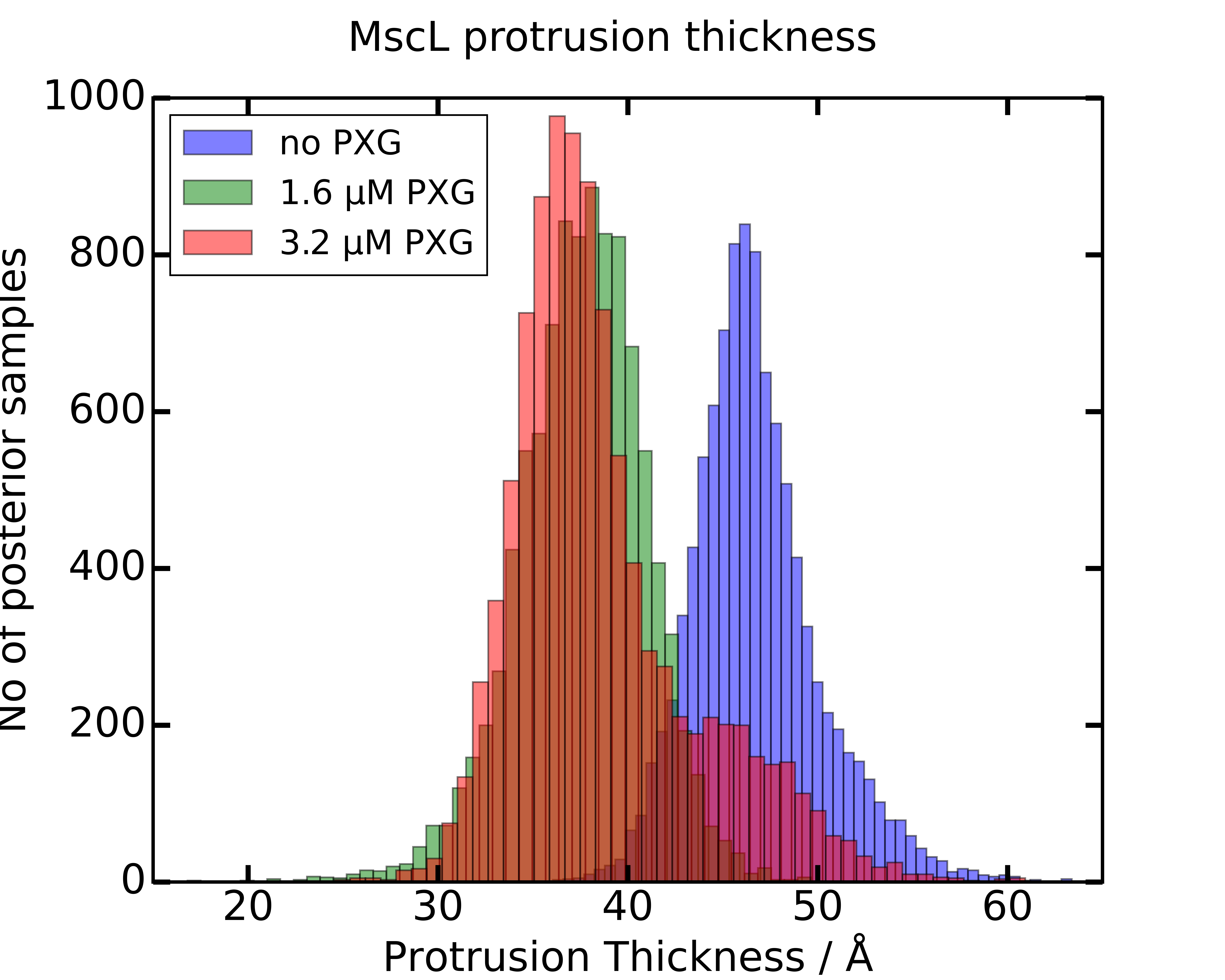}
  \caption{Posterior distributions of the MscL protrusion thickness of the
tethered MscL containing bilayer (initial bilayer (blue), after the addition of 1.6~$\mu$M PXG (green) and after 3.2~$\mu$M PXG (red).}
  \label{fgr:ProtrusionPosteriors}
\end{figure}

The most significant parameter change in this context occurs for the MscL protrusion thickness.
The significant shift of the whole posterior
probability distribution for this parameter to lower values compared to the pristine bilayer (Fig.~\ref{fgr:ProtrusionPosteriors}), provides strong support that there has been a conformational change in the C-terminus region of MscL. This could be consistent with the channel having gated to its open state. As we demonstrate in the Supplementary Information (SI section 2.2), the increase in the bilayer coverage parameter (see Table~\ref{tbl:bil_PXG}) that accompanies this change is also a signature of channel opening. 

The change in the density distribution for MscL (red line on Figures~\ref{fgr:NR_MscL_bilayer}-\ref{fgr:NR_MscL_bilayer_3p2_PXG}) appears to be consistent with the changes proposed by Bavi and co-workers to occur in both the C- and N-termini regions of the protein that sit either side of the head group region of the outer leaflet.\cite{bavi2017structural,Bavi_Nterm} They suggest that the upper third of the C-terminal helices (corresponding to a length $\sim$6~\AA) close to the bilayer/sub-phase interface bend outwards, increasing the channel permeability to the antibiotic streptomycin.\cite{bavi2017structural} They also suggest that the displacement of the N-terminus into the transmembrane region transmits tension from the bilayer to the channel causing it to gate open.\cite{Bavi_Nterm}  

MscL has been shown to have tension-induced gating activity when reconstituted into vesicles of various lipid compositions, including gel phase DPPE: DPPC and DSPE: DSPC bilayers \cite{nomura2012differential} and fluid phase POPC bilayers\cite{bavi2017structural}. The comparatively large APM determined for both leaflets of our bilayer mean that we expect it to be in the fluid phase. 
The related AMP, magainin 2 has been shown to induce a 2-3\% increase in the area of giant unilamellar vesicles (GUVs) by insertion into the hydrophobic tail region of the outer leaflet.\cite{Karal15} In the case of the GUVs, the resulting increase in membrane tension causes a pore to open, whereas we suggest that for MscL-containing bilayers it causes the channel to gate open as described by Bavi~{\it et al.}.\cite{Bavi_Nterm} We comment that for lyso-PC, a different mechanism has been proposed, in which the lyso-PC alters the coupling between the channel and the bilayer.\cite{Mukherjee14} As we don't have direct information on the location of the pexiganan, we cannot be conclusive about the mechanism by which pexiganan induces the conformational change in the MscL embedded in the bilayer.

\begin{table*}
\caption{Comparison between key structural parameters of the tethered MscL containing lipid bilayer determined from NR measured before and after the addition of PXG at two concentrations (1.6 and 3.2 $\mu$M). Additional model parameters are displayed in Tables S2-S4.}
\label{tab:PXG_bil}
\renewcommand{\arraystretch}{1.5}
\begin{tabular} {@{}lccc@{}}
\toprule
\textbf{Model parameter} & \textbf{MscL bilayer} & \textbf{+1.6~$\mu$M PXG} & \textbf{+3.2~$\mu$M PXG} \\ \hline %\midrule
waters per lipid head           &  6$\pm$1          &  5$\pm$2      & 5$\pm$1  \\
lipid APM outer (\AA{}$^2$)     &  99$\pm$2         &  98$\pm$2     & 98$\pm$2    \\
lipid APM inner (\AA{}$^2$)     & 79$\pm$1          & 79$\pm$1      & 79$\pm$1    \\
bilayer coverage                & 0.66$\pm$0.02     & 0.73$\pm$0.02 & 0.71$\pm$0.02  \\ 
bilayer roughness (\AA{})       &  7$\pm$1          & 9$\pm$1       & 9$\pm$1 \\
protein coverage                & 0.14$\pm$0.01     & 0.09$\pm$0.01  & 0.11$\pm$0.01 \\
MscL protrusion thickness (\AA{})  & 46$\pm$3          & 38$\pm$3      & 38$\pm$5 \\ 
\hline%\bottomrule
\label{tbl:bil_PXG}
\end{tabular}
\renewcommand{\arraystretch}{1}
\end{table*}

%%%%%%%%%%%%%%%%%%%%%%%%%%%%
\section{Conclusions}
%%%%%%%%%%%%%%%%%%%%%%%%%%%%

We have used a cell-free and detergent-free protocol to express the mechanosensitive ion channel of large conductance (MscL) into lipid vesicles. Characterization of these vesicles using SANS showed that the proteins were embedded in the lipid bilayer as single channels in contrast to the clusters found by a previous study which produced the protein by bacterial overexpression and used detergent. The roughness observed for the protein-containing vesicles suggests that the channel is oriented with the C-terminus domain outside the vesicle. An increase in the overall radius of gyration of the embedded proteins, indicating a change in conformation,  was observed upon the addition of the antimicrobial peptide PXG and Lyso-PC, a lipid with antimicrobial properties that is known to gate the ion-channel open.  We then used these protein-containing vesicles to produce planar, polymer-tethered, bilayers containing MscL. Using polarised neutron reflectometry (PNR) we were able to determine that the main contribution to the conformational change of the protein caused by PXG was a decrease in the thickness of the C-terminus protrusion pointing out from the membrane into the solvent. This decrease is consistent with existing models for channel opening. The areas per lipid molecule for the inner (78~\AA$^2$) and outer (99~\AA$^2$) leaflets of the the protein-containing bilayer are larger than typical for protein-free bilayers, suggesting that the bilayer may be under tension. The interaction with PXG also results in an increase in the fractional bilayer coverage, which  can be explained if the channel occludes a greater area, which would also be consistent with the PXG-induced conformational change corresponding to the channel opening.  The key features that enabled the observation of the conformational change are: the flexibility of the tethered bilayer, both in the plane and perpendicular to the plane, afforded by the area per molecule and the polymer tether respectively; the absence of channel clustering and detergent, which can both alter the probability of channel opening; and the sensitivity of SANS and PNR to resolve changes in the interfacial scattering length density. PNR is particularly sensitive to changes in scattering length density that occur perpendicular to the interface, which allows us to associate the main contribution to the conformational change with the C-terminus protrusion. 
Our structural investigations suggest that the conformational change could be consistent with the channel opening. As this could have important implications for the efficacy of antimicrobial peptides, an important next step would be to investigate whether PXG induces an opening of the channel using patch clamp ion channel recording. As there remains some debate as to the mechanism by which amphipathic molecules gate MscL open, a Molecular Dynamics investigation of PXG interacting with an MscL-containing bilayer would be useful and could help guide the design of antimicrobial peptides for therapeutic purposes.


%%%%%%%%%%%%%%%%%%%%%%%%%%%%
\section{Acknowledgments}
%%%%%%%%%%%%%%%%%%%%%%%%%%%%
We thank the ISIS Neutron Source for beam time in experiments RB1920647\cite{POLREF_rb1920647} and RB1820534\cite{POLREF_rb1820534} on PolRef and RB1910570\cite{SANS2D_rb1910570} and RB1820511\cite{SANS2D_rb1820511} on SANS2D. We also thank the CNST NanoFab at NIST, Gaithersburg, US for the provision of metal coatings for our substrates. We thank Boris Martinac and Paul Rohde, Victor Chang Cardiac Research Institute, for supplying the plasmid used in the cell-free expression.
SA was supported by the EPSRC CDT on “Soft and Functional Interfaces”
(EP/L015536/1) and an ISIS Facility Development Studentship.
For the purpose of open access, the author has applied a Creative Commons Attribution (CC BY) licence to any Author Accepted Manuscript version arising from this submission.

%%%%%%%%
\bibliographystyle{elsarticle-num}
\bibliography{Ayscough25}

\end{document}

%%%%%%%%%%%%%%%%%%%%%%%%%%%%%%%%%%%%%%%%%%%%%%%%%%%%%%%%%%%%%%%%%%%%%
%% Place any additional macros here.  Please use \newcommand* where
%% possible, and avoid layout-changing macros (which are not used
%% when typesetting).
%%%%%%%%%%%%%%%%%%%%%%%%%%%%%%%%%%%%%%%%%%%%%%%%%%%%%%%%%%%%%%%%%%%%%
\newcommand*\mycommand[1]{\texttt{\emph{#1}}}
\let\endtitlepage\relax

\renewcommand{\figurename}{Fig.}
\renewcommand{\thefigure}{S\arabic{figure}}

\renewcommand{\tablename}{Table}
\renewcommand{\thetable}{S\arabic{table}}

\renewcommand{\thesection}{\arabic{section}}

\author{Sophie E. Ayscough}
\affiliation{School of Physics \& Astronomy, James Clerk Maxwell Building, University of Edinburgh, Edinburgh, EH9 3FD, UK}
\affiliation{Lund University, Lund 22100, Sweden}
\author{Maximilian W. A. Skoda}
\email{maximilian.skoda@stfc.ac.uk}
\affiliation{ISIS Neutron and Muon Source, Rutherford Appleton Laboratory, Harwell Campus, Chilton, OX11 0QX, UK}
\author{James Doutch}
\affiliation{ISIS Neutron and Muon Source, Rutherford Appleton Laboratory, Harwell Campus, Chilton, OX11 0QX, UK}
\author{Andrew Caruana}
\affiliation{ISIS Neutron and Muon Source, Rutherford Appleton Laboratory, Harwell Campus, Chilton, OX11 0QX, UK}
\author{Christy Kinane}
\affiliation{ISIS Neutron and Muon Source, Rutherford Appleton Laboratory, Harwell Campus, Chilton, OX11 0QX, UK}
\author{Luke Clifton}
\affiliation{ISIS Neutron and Muon Source, Rutherford Appleton Laboratory, Harwell Campus, Chilton, OX11 0QX, UK}
\author{Simon Titmuss}
\email{simon.titmuss@ed.ac.uk}
\affiliation{School of Physics \& Astronomy, James Clerk Maxwell Building, University of Edinburgh, Edinburgh, EH9 3FD, UK}

\title{Supplementary Information: Neutron Reflectometry Reveals Conformational Changes in a Mechanosensitive  Protein Induced by an Antimicrobial Peptide in Tethered Lipid Bilayers}

\maketitle

\section{Quartz-crystal microbalance with dissipation (QCM-D)}\label{sec:SI_QCM}
A typical QCM-D protocol followed the following steps:
\begin{enumerate}
\item Buffer solution (10 mL) was passed through each of the 
flow cells and the measurement started with the sensors in HEPES buffer.

\item After testing that the frequency is stable for 30 min with only clean
buffer in the cell, i.e. there is no desorption of tether occurring, adsorption
of contaminants or bubbles in the system, then a solution of vesicles was
passed through the cells at 1 mL/min using the peristaltic pump. 10 ml of
vesicle solution is passed through each QCM cell.

\item After 1~h, 300~mM NaCl HEPES buffer was passed through the cells to remove excess vesicles.
\item This was immediately followed by a salt rupture step, pushing 5~mL of lower
salt concentration 150~mM NaCl HEPES buffer through the QCM cell.
\end{enumerate}

Since tethered bilayers are viscoelastic it was inappropriate to analyse the results
quantitatively using the Sauerbrey equation. The system has high water content
in the tether layer and also in any tethered vesicles. High water content,
 such as when the vesicles remain intact on the tether layer, increases elasticity,
which results in a high $-\Delta D/\Delta F$ value.~\cite{INCI2015} When the vesicles rupture, a
more rigid layer is formed resulting in the $-\Delta D/\Delta F$ ratio decreasing. Using
$-\Delta D/\Delta F$ to assess vesicle rupture has been employed by several groups.~\cite{Hopfner08,INCI2015}
Though the added complication of a long tether layer has been shown to distort
values. A study into the impact of PEG2000-DSPE as a spacer molecule for
membrane tethering found $-\Delta D/\Delta F$ values between 1.8 and 5 for PEG tethered
bilayers.~\cite{INCI2015} The value is dependent on the density of tethering as well as the
bilayer structure. The QCM study by Inci \textit{et al.} suggested that in their system,
individual vesicle rupture occurred and therefore a decrease in dissipation and
increase in frequency was not observed.~\cite{INCI2015}

%%%%%%%%%%%%%%%%%%%%%%%%%%%%%%%%%%%%%%%%%%%%%%%%%%%
\section{Neutron reflectivity data analysis}
%%%%%%%%%%%%%%%%%%%%%%%%%%%%%%%%%%%%%%%%%%%%%%%%%%%
While the layers from silicon oxide to poly(ethylene glycol) (PEG2000) were parameterised as simple slabs, the DSPE and lipid bilayer were parameterised in terms of area per molecule (APM) and water molecules per lipid head (WPLH) using literature values for molecular volumes and atomic scattering lengths.\cite{Armen1998, Dabkowska2012, Pan2012, Sears1992} This has the advantage of being able to directly obtain a physically meaningful quantity (APM) and it provides the ability to apply physical constraints, such as matching numbers of head groups and tails. Values for the lipid head and tail group volumes were taken from the RefNX database\cite{Nelson2019, CHIU1999, Pabst2007-lm}

The SLDs of the silicon and silicon oxide were fixed at literature values. As the SLD of the permalloy layer has a magnetic component, that is dependent on the strength of the applied static field and the exact composition of the film, its value was determined by fitting the data measured from the substrates functionalized  with the tether  and then fixed. The thicknesses and interfacial widths for gold, permalloy and silicon oxide layers were also determined from the best fit to the tether data and then fixed for the subsequent data sets. The best fit parameter values are shown in Table~\ref{tbl:metals}.

When treating the MscL containing bilayer, the APM and WPLH parameters for each of the two bilayer leaflets were treated independently.
The composition of the the bilayer was assumed to be 3:1 POPC:POPG lipids despite the lower leaflet containing a high coverage of DSPE. The difference in the lipid tail and head volume between POPC, POPG and DSPE is sufficiently small that this will have negligible impact on the fitted parameters. To incorporate protein into the model, we fit the volume fraction of protein in the transmembrane region (protein coverage parameter) and a protrusion thickness. The thickness of the transmembrane domain is constrained to that of of the lipid bilayer, as determined by the APMs and WPLH for the inner and outer leaflets. The number of protrusions is constrained to equal the number of  transmembrane domains, as determined by the protein coverage.
Although models were tested with the protrusion oriented on the tether side, the subphase side and on both sides of the bilayer, favourable fits were only obtained when the protrusion was oriented outwards, towards the subphase. This was quantified by evaluating the log-evidence,logz, as: 3744, in the absence of protrusion, 3792, with protrusion directed towards substrate, and 3876, with the protrusion directed towards the subphase; a higher logz indicates a more likely model.

PXG is not explicitly included in the model.  Our previous studies indicated a low peptide to lipid ratio,\cite{mckinley2017} meaning that the small contribution to the scattering length density of the layer does not warrant the concomitant increase in the complexity of the model.

A differential evolution minimization is used to adjust the fit parameters to reduce the differences between the model reflectivity and the data. In all cases the simplest possible model (i.e. fewest layers), which adequately described the data, was selected. 
The uncertainties on the best fit parameters were assessed within a Bayesian framework, in which the posterior parameter probability distributions were sampled using a MCMC approach within RefNX.
For the sampling, we used 800 “burn in” points followed by 4000 samples with a thinning of 100. The resulting plots contain fits and corresponding real space structure of the sample layer system, as well as 300 samples from the posterior distributions (shown as shaded lines/regions).

%%%%%%%%%%%%%%%%%%%%%%%%%%%%%%%%%%%%%%%%%%%%%%%%%%%
\subsection{Neutron reflectivity fit parameters}
\begin{figure}[!htp]
\includegraphics[width=0.6\textwidth]{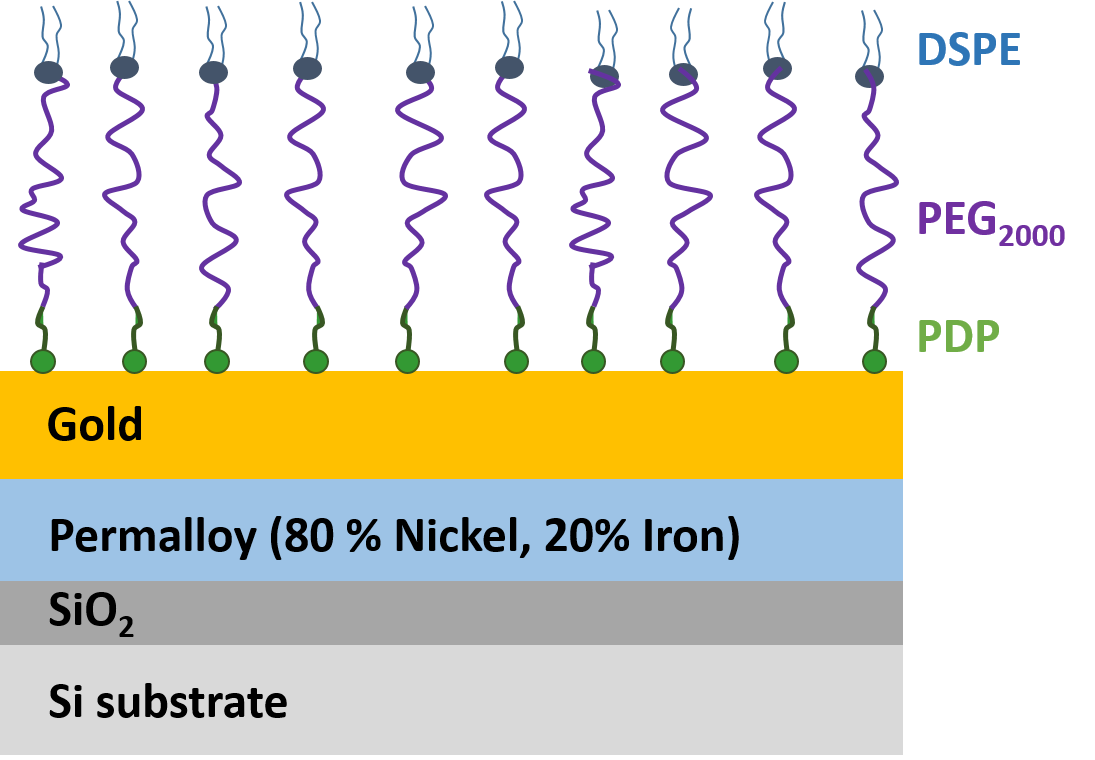}
  \caption{Schematic representation (not to scale) of the layer structure.}
  \label{fgr:layer_structure}
\end{figure}

%%%%%%%%%%%%%%%%%%%%%%%%%%%%%%%%%%%%%%%%%%%%%%%%%%%
A complete implementation of the fitting model, as used in the RefNX software can be found in the section~\ref{listing} (bilayer defintion is in listing lines 220-320).
The model consisted of the following layers, as determined by the sample architecture:
Si substrate, oxide, permalloy (Py), gold (Au), PDP and PEG (Fig.~\ref{fgr:layer_structure}).

In addition, a bilayer was constructed based on an area per molecule model and employing literature values for the lipids used (POPC and POPG)\cite{Nelson2019, CHIU1999, Pabst2007-lm}. The area per molecule description was then converted into a slab (layer) structure for fitting by the RefNX software. The tables below contain a list of the fitted model parameters and the prior fitting bounds for each. After the co-refinement of the tether only data set, the values for the inorganic layers (oxide, permalloy and gold) were kept fixed throughout (Table~\ref{tbl:metals}).

\begin{table}[!htp]
 \caption{Fit parameters for the inorganic layers.}
\renewcommand{\arraystretch}{1.5}
\begin{tabular}{l c c c} %{@{}lccc@{}}
\toprule
\textbf{Model parameter} & \textbf{Best fit value} & \textbf{Standard error} & \textbf{Prior bounds} \\ \midrule
        oxide thickness (\AA{})& 15 & 2 & [10, 20] \\ 
        substrate/oxide roughness (\AA{})& 6 & 1 & [5, 10] \\ 
        permalloy thickness (\AA{}) & 130.4 & 0.3 & [110, 180] \\ 
        permalloy SLD up ($\times10^{-6}$\AA{}$^{-2}$)& 10.00 & 0.02 & [8, 12] \\ 
        permalloy SLD down ($\times10^{-6}$\AA{}$^{-2}$)& 7.20 & 0.01 & [5, 8] \\ 
        permalloy roughness (\AA{})& 6 & 1 & [5, 20] \\ 
        Au thickness (\AA{})& 146.5 & 0.3 & [120, 200] \\ 
        Au SLD ($\times10^{-6}$\AA{}$^{-2}$)& 4.2 & 0.01 & [4.2, 4.6] \\ 
        Au roughness (\AA{})& 10.0 & 0.3 & [3, 15] \\
        \bottomrule
    \end{tabular}
    \label{tbl:metals}
\end{table}

\begin{table}[!htp]
 \caption{Fit parameters MscL containing bilayer before PXG.}
\renewcommand{\arraystretch}{1.5}
\begin{tabular}{l c c c} %{@{}lccc@{}}
\toprule
\textbf{Model parameter} & \textbf{Best fit value} & \textbf{Standard error} & \textbf{Prior bounds} \\ \midrule

        PDP thickness (\AA{}) & 8.6 & 2.4 & [5, 15] \\
        PDP hydration$^{\ast}$ & 0.23 & 0.04 & [0.2, 0.6] \\
        PEG thickness (\AA{}) & 79 & 3 & [40, 100] \\ 
        PEG roughness (\AA{})& 7 & 1 & [5, 25] \\ 
        PEG hydration$^{\ast}$ & 0.480 & 0.006 & [0.2, 1.0] \\ 
        POPC/POPG tail volume (\AA{}$^3$) & 944.0 & N/A & fixed \\
        POPC head volume (\AA{}$^3$) & 319.0 & N/A & fixed \\
        POPG head volume (\AA{}$^3$) & 257.0 & N/A &fixed \\
        Waters Per Lipid Head Group (WPLH) & 6 & 1 & [5.0, 30.0]  \\ 
        lipid APM outer  (\AA{}$^2$) & 99 & 2 & [40.0, 100.0]  \\
       lipid APM inner (\AA{}$^2$)  & 79 & 1 & [50.0, 80.0]  \\ 
        bilayer coverage & 0.66 & 0.02 & [0.0, 1.0]  \\ 
        MscL protrusion thickness (\AA{}) & 46 & 3 & [0.0, 65.0]  \\ 
        protein coverage (\%) & 13.7 & 1.4 & [0, 30]  \\ 
        bilayer/solvent roughness (\AA{})&	5.2 &	3.5 & 	[2.0, 15.0] \\
        common scale & 1.000 & 0.002 & [0.9, 1.1]  \\ 
        background ($\times10^{-6}$) & 8.1 & 0.4 & [0.1, 10]  \\ 
        D$_2$O SLD ($\times10^{-6}$\AA{}$^{-2}$) & 6.22 & 0.01 & [6.1, 6.36] \\
        GMW SLD ($\times10^{-6}$\AA{}$^{-2}$)& 4.56 & 0.01 & [4.1, 4.7] \\
        H$_2$O SLD ($\times10^{-6}$\AA{}$^{-2}$) & -0.55 & 0.01 & [-0.56, 0.0] \\
        \bottomrule
        \multicolumn{3}{l}{\textsuperscript{*}\footnotesize{volume fraction of solvent in the layer}}
    \end{tabular}
\end{table}

\begin{table}[!htp]
\caption{Fit parameters bilayer with 1.6 $\mu$M PXG.}
\renewcommand{\arraystretch}{1.5}
\begin{tabular}{l c c c} %{@{}lccc@{}}
\toprule
\textbf{Model parameter} & \textbf{Best fit value} & \textbf{Standard error} & \textbf{Prior bounds} \\ \midrule
        PDP thickness (\AA{}) & 7.6 & 0.5 & [5, 15] \\ 
        PDP hydration$^{\ast}$ & 0.21 & 0.01 & [0.2, 0.6] \\ 
        PEG thickness (\AA{}) & 79 & 1 & [40, 100] \\ 
        PEG roughness (\AA{}) & 9.4 & 0.5 & [5, 25] \\ 
        PEG hydration$^{\ast}$ & 0.5 & 0.0 & [0.2, 1.0] \\ 
         Waters Per Lipid Head Group (WPLH) & 5 & 2 & [5, 30] \\ 
        lipid APM outer (\AA{}$^2$) & 98 & 2 & [40, 120] \\ 
        lipid APM inner (\AA{}$^2$) & 79 & 1 & [50, 120] \\ 
        bilayer coverage & 0.73 & 0.02 & [0.0, 1.0] \\ 
        MscL protrusion thickness (\AA{}) & 38 & 3 & [0, 65] \\ 
        protein coverage (\%) & 9.0 & 1.0 & [0, 30] \\ 
        bilayer/solvent roughness (\AA{}) & 6 & 3 & [2, 15] \\ 
        common scale & 1.0 & 0.0 & [0.9, 1.1] \\ 
        background ($\times10^{-6}$)& 8.3 & 0.4 & [0.1, 10] \\ 
        D$_2$O SLD ($\times10^{-6}$\AA{}$^{-2}$)& 6.31 & 0.01 & [6.1, 6.36] \\         
        GMW SLD ($\times10^{-6}$\AA{}$^{-2}$) & 4.46 & 0.01 & [4.1, 4.7] \\
        H$_2$O SLD ($\times10^{-6}$\AA{}$^{-2}$) & -0.556 & 0.005 & [-0.56, 0.0] \\

\bottomrule
        \multicolumn{3}{l}{\textsuperscript{*}\footnotesize{volume fraction of solvent in the layer}}
\label{tbl:bil_1p6_PXG}
\end{tabular}
\renewcommand{\arraystretch}{1}
\end{table}

\begin{table}[p]
\caption{Fit parameters bilayer with 3.2 $\mu$M PXG.}
\renewcommand{\arraystretch}{1.5}
\begin{tabular}{l c c c} %{@{}lccc@{}}
\toprule
\textbf{Model parameter} & \textbf{Best fit value} & \textbf{Standard error} & \textbf{Prior bounds} \\ \midrule
        PDP thickness (\AA{}) & 7.2 & 0.5 & [5, 15] \\ 
        PDP hydration$^{\ast}$ & 0.21 & 0.01 & [0.2, 0.6] \\ 
        PEG thickness (\AA{}) & 80 & 1 & [40, 100] \\ 
        PEG roughness (\AA{}) & 9.2 & 0.6 & [5, 25] \\ 
        PEG hydration$^{\ast}$ & 0.514 & 0.003 & [0.2, 1.0] \\ 
        Waters Per Lipid Head Group (WPLH) & 5 & 1 & [5, 30] \\ 
        lipid APM outer (\AA{}$^2$) & 98 & 2 & [40, 120] \\ 
        lipid APM inner (\AA{}$^2$) & 79 & 1 & [50, 120] \\ 
        bilayer coverage & 0.71 & 0.02 & [0.0, 1.0] \\ 
        MscL protrusion thickness (\AA{}) & 38 & 5 & [0, 65] \\ 
        protein coverage (\%)& 11 & 1 & [0, 30] \\ 
        bilayer/solvent roughness (\AA{}) & 4 & 2 & [2, 15] \\ 
        common scale & 1.0 & 0.0 & [0.9, 1.1] \\ 
        background ($\times10^{-6}$)& 6.7 & 0.8 & [0.1, 10] \\ 
        D$_2$O SLD ($\times10^{-6}$\AA{}$^{-2}$)& 6.30 & 0.01 & [6.1, 6.36] \\         
        GMW SLD ($\times10^{-6}$\AA{}$^{-2}$) & 4.46 & 0.01 & [4.1, 4.7] \\
        H$_2$O SLD ($\times10^{-6}$\AA{}$^{-2}$) & -0.56 & 0.01 & [-0.56, 0.0] \\
\bottomrule
        \multicolumn{3}{l}{\textsuperscript{*}\footnotesize{volume fraction of solvent in the layer}}
\label{tbl:bil_3p2_PXG}
\end{tabular}
\renewcommand{\arraystretch}{1}
\end{table}
\subsection{Justification of observed variation in bilayer coverage}
The results given in Tables 2 and S2-S4 indicate that the best fit bilayer coverage increases from $0.66\pm 0.02$ to $0.74\pm 0.02$ following the addition of 1.6~$\mu$M pexiganan. This parameter represents the fraction, $f_B$, of the neutron beam footprint area, $A_{TOT}$ that is covered by protein containing bilayer, with the complement being buffer. As the following simple calculation illustrates, the observed increase in this coverage is a consequence of the open channel, $A_{P}^\prime$, occluding a greater area than the closed channel, $A_P$. 
Using the bilayer ($f_B$) and protein ($f_P$) coverages given in Table~S2 for the bilayer before gating implies the total covered area $A_C=0.66A_{TOT}$, the area covered by lipids $A_L=(1-f_P)f_BA_{TOT} = 0.57A_{TOT}$ and that by protein, $A_P=f_Pf_BA_{TOT}=0.092A_{TOT}$. Using the simplified geometry given in Figure~\ref{MscLGating}, the area occluded by each protein channel after opening will increase by a factor of $(35^2-15^2)/25^2=1.6$, such that the total area now occluded by protein is $A_{P}^\prime = 0.15A_{TOT}$. As the APM of the lipids does not change, the area covered by lipids does not change. This means that the total covered area increases to $A_{P}^\prime+A_L = 0.72A_{TOT}$ corresponding to a bilayer coverage of $f_{B}^\prime=0.72$, which is  within the uncertainty given for the best value in Table~S3.

As the protein coverage gives the volume fraction of protein in the transmembrane domain, the increase in the total area of the protein-containing bilayer when the channels gate to open, means that the protein coverage is expected to decrease, in line with the best values given in Tables 2 and S2-S4.
%\FloatBarrier
\section{Error analysis posterior distributions}

\begin{figure}[p]
\includegraphics[width=\textwidth]{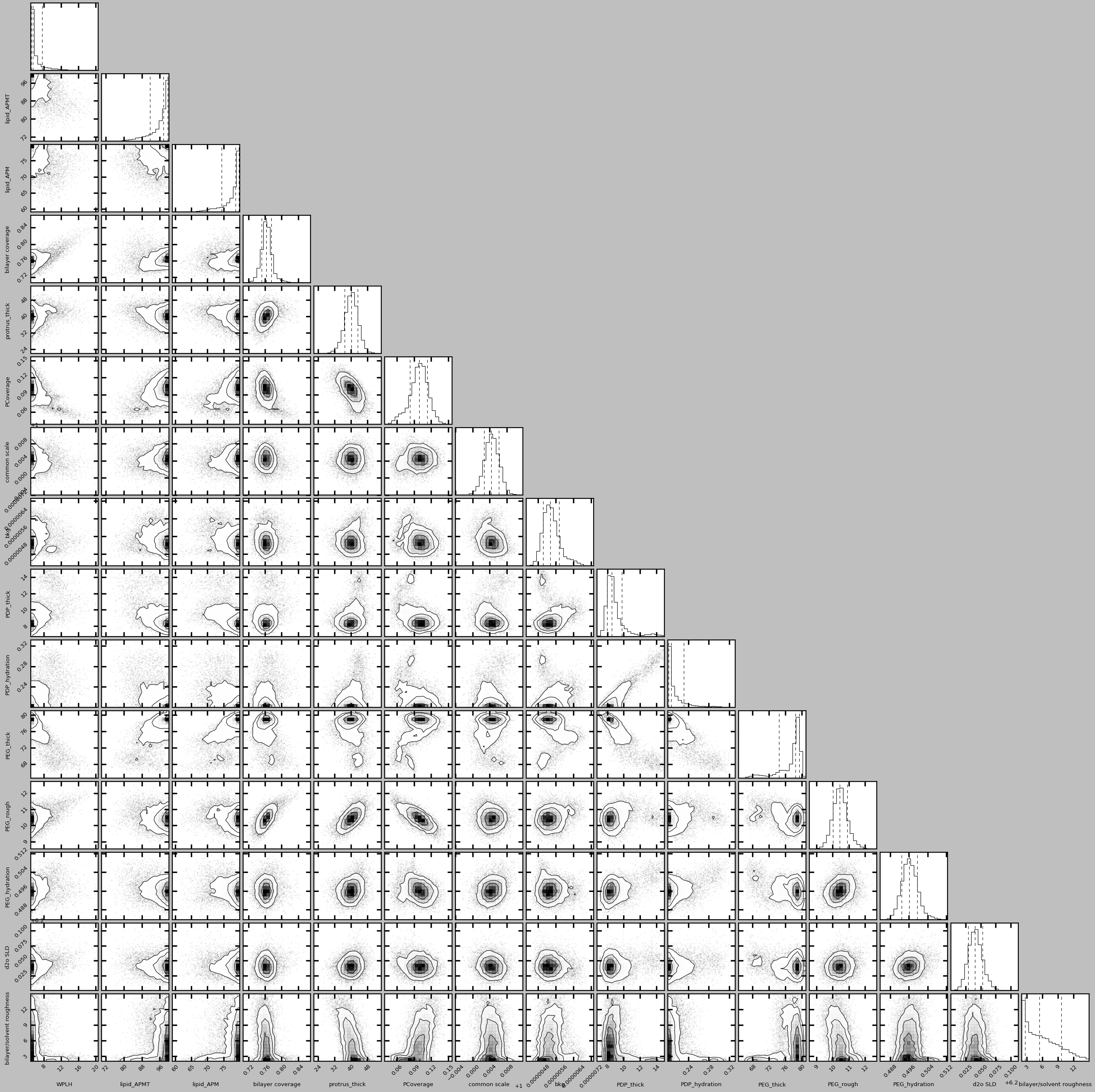}
  \caption{Corner plot with posterior distributions for the fitting parameters of the MscL containing bilayer without PXG.}
  \label{fgr:bilayer_only_corner}
\end{figure}

\begin{figure}[p]
\includegraphics[width=\textwidth]{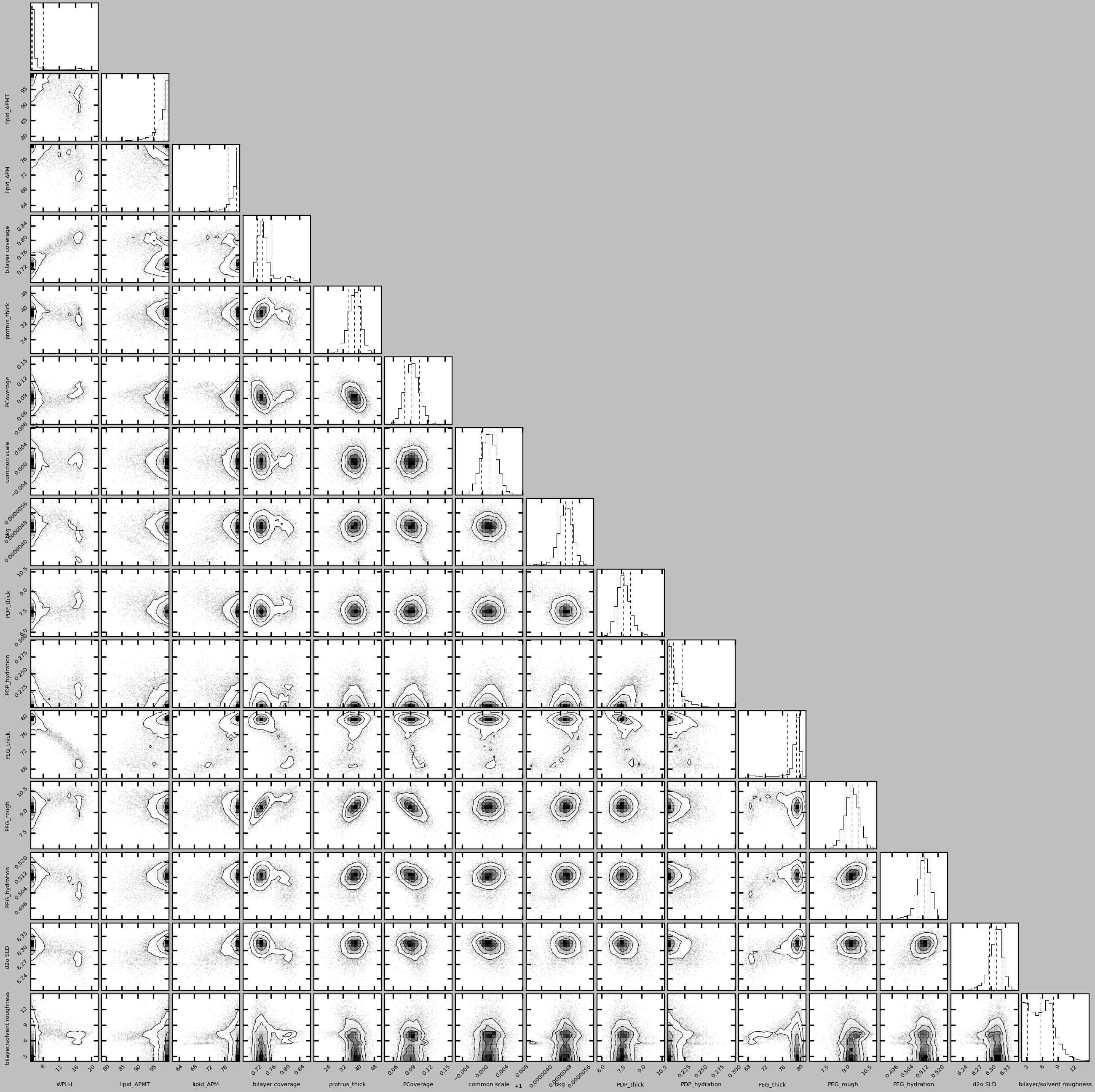}
  \caption{Corner plot with posterior distributions for the fitting parameters of the MscL containing bilayer with 1.6~$\mu$M PXG.}
  \label{fgr:bilayer_1p6_PXG_corner}
\end{figure}

\begin{figure}[p]
\includegraphics[width=\textwidth]{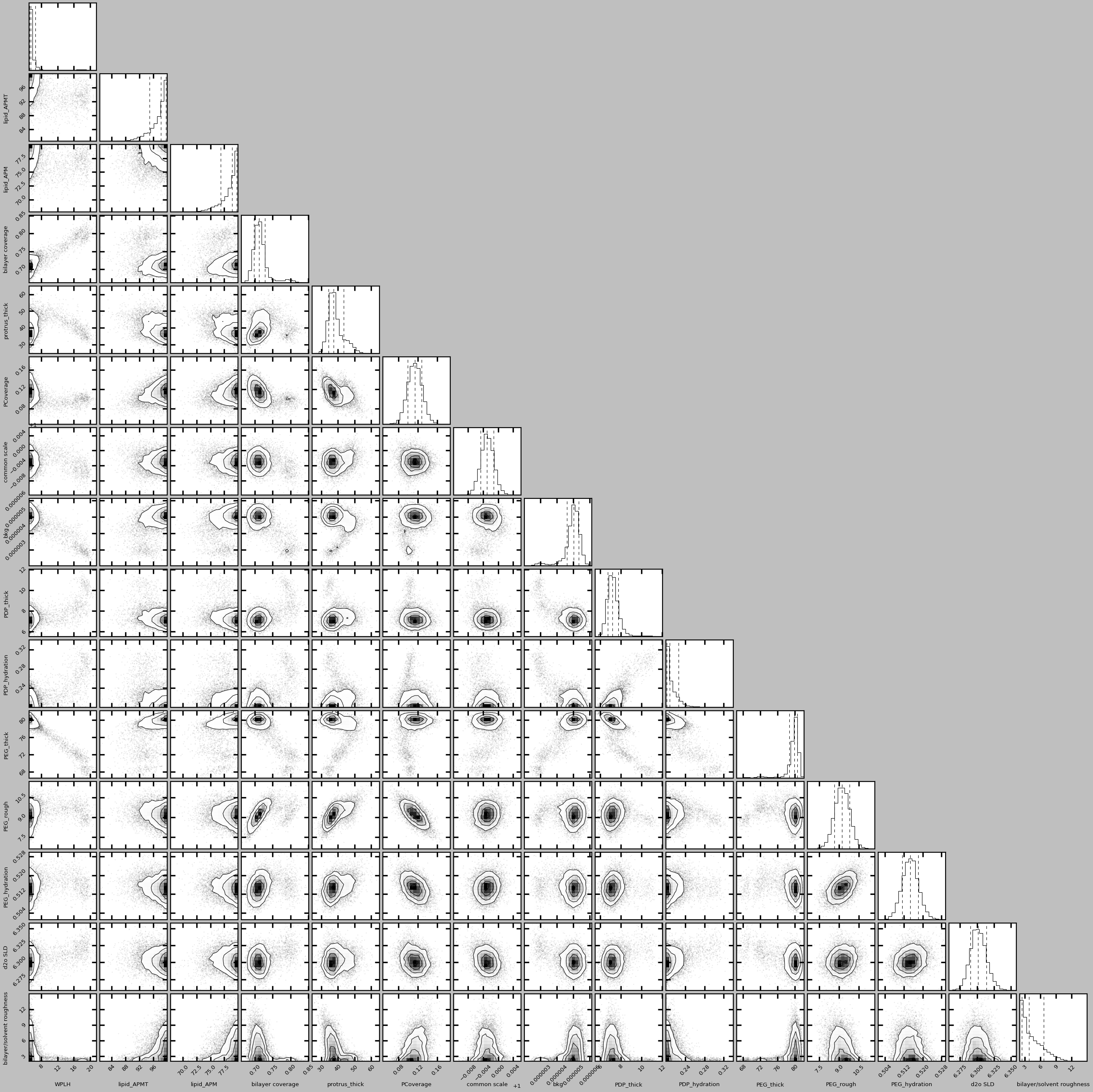}
  \caption{Corner plot with posterior distributions for the fitting parameters of the MscL containing bilayer with 3.2~$\mu$M PXG.}
  \label{fgr:bilayer_3p2_PXG_corner}
\end{figure}

RefNX uses a Bayesian approach employing a Markov-chain Monte Carlo algorithm to investigate the posterior probability distribution of the model parameters. The Bayesian approach is useful for examining parameter covariances. The figures below (Figs.~\ref{fgr:bilayer_only_corner}, \ref{fgr:bilayer_1p6_PXG_corner} and \ref{fgr:bilayer_3p2_PXG_corner}) show the corner plots for the conditions without PXG and with 1.6 and 3.2~$\mu$M PXG respectively.
The corner plots illustrate the correlations between pairs of parameters and also show the final posterior distribution for each parameter. The standard error associated with each parameter is half the width of the (15,85) percentile of the corresponding posterior distribution.

\section{Cell-free protein expression}\label{cell-free}
First a concentrated solution (18~mg/mL) of 3:1 POPC:POPG lipids was
prepared. 13.5~mg of POPC and 4.5~mg of POPC were dissolved in the minimum
amount of chloroform and the chloroform removed under a stream of nitrogen
to produce a dried lipid film. Then 0.976 mL of BiotechRabbit reconstitution
buffer was added to the lipid film and the solution tip-sonicated for 30 minutes at room temperature. The vesicle mixture was then ready to be added to the
final reaction solution as described below.

All components of the RTS500 BiotechRabbit kit were stored at -20$^\circ$C until the
day of expression. The Dithiorithretol (DTT) solution, reconstitution buffer and
MscL plasmid were thawed at room temperature whilst all other components of
the RTS kit were thawed on ice (E.coli lysate, feeding mix, reaction mix). After
thawing, components were reconstituted by adding in an appropriate volume of
reconstitution buffer and lightly rolling the reaction vessel (no shaking) to avoid
shearing of delicate biological components. After reconstitution, components
were kept on ice until they were combined into the final reaction vessel. The
E.coli lysate was reconstituted in buffer (0.2~mL) and vesicle solution (0.32~mL).
The reaction mix was reconstituted in buffer (0.22~mL), the amino acid mix was
reconstituted in buffer (3~mL) and the methionine was reconstituted in buffer (1.8~mL). 
Finally the lyophilised feed mix was reconstituted with buffer (8.1~mL).
After reconstitution, all components were then combined into a reaction solution
and a feeding mix was placed into the reaction vessel (supplied by
BiotechRabbit).\cite{biotechrabbit2023} Splitting the solutions into a reaction mix and feeding mix this way, allowed for a higher yield to be produced. In the reaction compartment
 high concentrations of the critical components are present, whilst small components
such as amino acids can be exchanged into the reaction compartment and waste
products can diffuse out. The feeding mix was prepared by adding lyophilised
feed mix (8.1~mL), amino acid solution (2.6~mL), methionine solution (0.3~mL) and
DTT solution (0.3~mL). The reaction solution was prepared by adding reaction
mix (0.22~mL), reconstituted \textit{E. coli} mix (0.52~mL), amino acid solution (0.27~mL)
and methionine (30~$\mu$L). Finally a solution of the expression plasmid 
(10~$\mu$L at 464~$\mu$g/mL) was added to reaction mix and the reaction container was assembled and placed in a shaker incubator at 30~$^\circ$C for 23 hours.

After 23 hours of incubation, the reaction mix was removed from the reaction container and centrifuged at 4~$^\circ$C for 1~h to produce pellets, the supernatant removed
and the pellet resuspended in 4-(2-hydroxyethyl)-1-piperazineethanesulfonic
acid (HEPES) buffer (1~mL, 20~mM, pH 7, 100~mM KCl). The 
solution of protein containing vesicles was then characterised as described in \ref{CF1} and \ref{CF2} below, stored at 4~$^\circ$C until use, and used within a week. Storage at 4~$^\circ$C is recommended in various protocols that subsequently use ion-channel recording to monitor channel activity, \cite{Najem15,Martinac_Weissig} and we do not observe any sign of aggregation from our SANS measurements. 

\subsection{Gel Electrophoresis}\label{CF1}
\begin{figure}[p]
\includegraphics[width=0.3\textwidth]{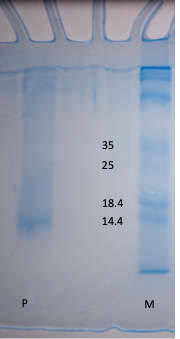}
  \caption{Photograph of run SDS-PAGE 12\% acrylamide gel of cell free protein expression of MscL after SDS treatment to remove lipids (P) and the molecular marker EZ-marker BP3600 (M), stained with Coomassie Brilliant Blue. The different weights of the marker bands are labelled for clarity. The gel suggests a protein has been expressed with a weight of about 14 kDa, which is in agreement of the monomeric unit of MscL which should be 14.2 kDa. }
  \label{GelElectrophoresis}
\end{figure}

Sodium Dodecyl Sulphate Poly Acrylamide Gel Electrophoresis (SDS-PAGE) was run using 12\% acrylamide gels. Aliquots (10 µl) were taken from each protein expression and mixed with aliquots of Sigma Aldrich SDS-PAGE staining solution (10 $\mu$L) and heated at 60°C (1 hour) and cooled to room temperature before loading onto the gel. Lipids were not removed from the protein sample before analysis which may have led to some blurring of the protein bands. After the
samples had run on the gels (approximately 1.5 hours), the gels were removed from the glass supports and rinsed multiple times with Milli-Q water. The gels were left to soak in Milli-Q water for 1 hour, the water was then replaced and left again for an hour to remove SDS from the gel. Gels were stained with ThemoFisher Coomassie safe stain by covering the gels in the solution for an hour followed by rinsing and soaking (1 hour) with Milli-Q water. A photo was then taken of the gel and is shown in Figure~\ref{GelElectrophoresis}.

\subsection{Protein concentration determination} \label{CF2}
The BCA (Bicinchroninic Acid) method of protein quantification was used following the protocol from Takeda, M. \textit{et al.}.\cite{Takeda2012} A Thermo Fisher Scientific Pierce BCA protein assay kit was used which contained Bovine Serum Albumin (BSA) standards for the expressed protein to be quantified against.
The solubilisation buffer was prepared from Triton-X 100 (0.2\%), solubilised in HEPES-KOH buffer (20 mm, pH 7.4). An aliquot (60 $\mu$L) of the protein containing vesicle sample was made up to 1500 $\mu$L with solubilisation buffer, decanted into three separate tubes (500 $\mu$L). Cold acetone (1 mL) was added to each of the tubes, the tubes vortexed and incubated for 20 minutes at -20$^{\circ}$C. The tubes were then centrifuged at room temperature (10 000g, 10 minutes), the supernatant discarded and pellets incubated for 30 minutes at room temperature to allow any remaining acetone to evaporate. Solubilisation buffer (500 $\mu$L) was added to re-suspend the pellets and the tubes vortexed. BCA working reagent (500 ul) was added and the tubes incubated at 60 $^{\circ}$C for 1 hour. Samples were allowed to cool for 10 minutes before measurement. Bovine Serum Albumin (BSA) standards were prepared over a concentration
range of 0.5-20 $\mu$g in the same solubilisation buffer as the protein containing vesicle samples to produce a calibration against which we could quantify our protein. Each standard (500 $\mu$L) was prepared in the same way as the protein containing vesicle samples by adding BCA working reagent (500 $\mu$L) and heating for an hour at 60$^{\circ}$C before cooling to room temperature for 10 minutes. A spectrophotometer was then used to measure absorbance at 562 nm.

\section{Small angle neutron scattering analysis}
Hammouda \textit{et al.} proposed a generalized Guinier-Porod  approach to fitting the scattering from a mixture of arbitrary shapes or fractal structures for which it is difficult to build analytical models.\citep{hammouda2010new}

For each particulate type in the system (described as a level in the Irena implementation) the scattered intensity is described as the sum of two terms: 

\begin{align}
\begin{split}
I(Q) &=\frac{G}{Q^s}\exp \left(\frac{-Q^{2}R_g^{2}}{3-s}\right)\text{for} \ Q \ \leq \ Q_1 
\\
I(Q) &= \frac{D}{Q^d} \ \text{for} \ Q \ \geq  \ Q_1,
\end{split}
\end{align}
where, $Q$ is the scattering vector, $R_g$ is the radius of gyration of the particle contributing to that level, $s$ is the shape parameter, $d$ is the Porod exponent and $G$ and $D$ are respectively the Guinier and Porod region scale factors. For globular structures such as spheres $s=0$, whilst for rods and platelets $s=1$ and $s=2$, respectively. The Porod exponent provides information about the interface of the corresponding scattering object (Table~\ref{Porod}).

\begin{table*}[h]
\begin{center}

\caption{Table of Porod exponents and its relation to common scattering objects.}
\label{Porod}
\begin{tabular}{|p{4cm}|p{6cm}|}
\hline
Porod exponent, $d$ & Nature of scattering object\\
\hline
4 & Very smooth sphere\\
3 & Very rough object or collapsed polymer chains\\
2 & Gaussian polymer chain or 2-D structure(lamellae or platelets)\\
1 & Stiff rod or thin cylinder\\
\hline
\end{tabular}
\end{center}
\end{table*}

All fitting to Guinier-Porod models was carried out on Igor v.6.37 using the NCNS Irena SAS macro v.2.63.\citep{GPmodelirena}
The Irena Macro allows for multiple Guinier-Porod levels to be fitted where each level corresponds to a different scattering object.\citep{GPmodelirena} This is important for our system where in all measured contrasts there will be scattering from the protein (level=GP1) and from the overall protein-containing vesicle structure (level = GP2).

\subsection{Estimation of Radius of Gyration of MscL}\label{MscL_Rg}

The radius of gyration of a scattering object describes the second moment of the distribution of the mass of the object:

\begin{equation}
R_g^2 = \left(\sum_{i} m_ir_i^2 / \sum_i m_i \right) ,
\end{equation}

where $m_i$ is the mass of an atom and $r_i$ is the distance of an atom from the centre of mass. 
Applying  this to the pdb file of the 2OAR crystal structure of MscL, using a number of free online resources, \citep{Flitersforglobularprotein, van2005gromacs, chang1998structure, xie2013studies} yields a radius of gyration of 27.8~\AA{}.

\begin{figure}
\centering
  
\includegraphics[width=0.9\textwidth]{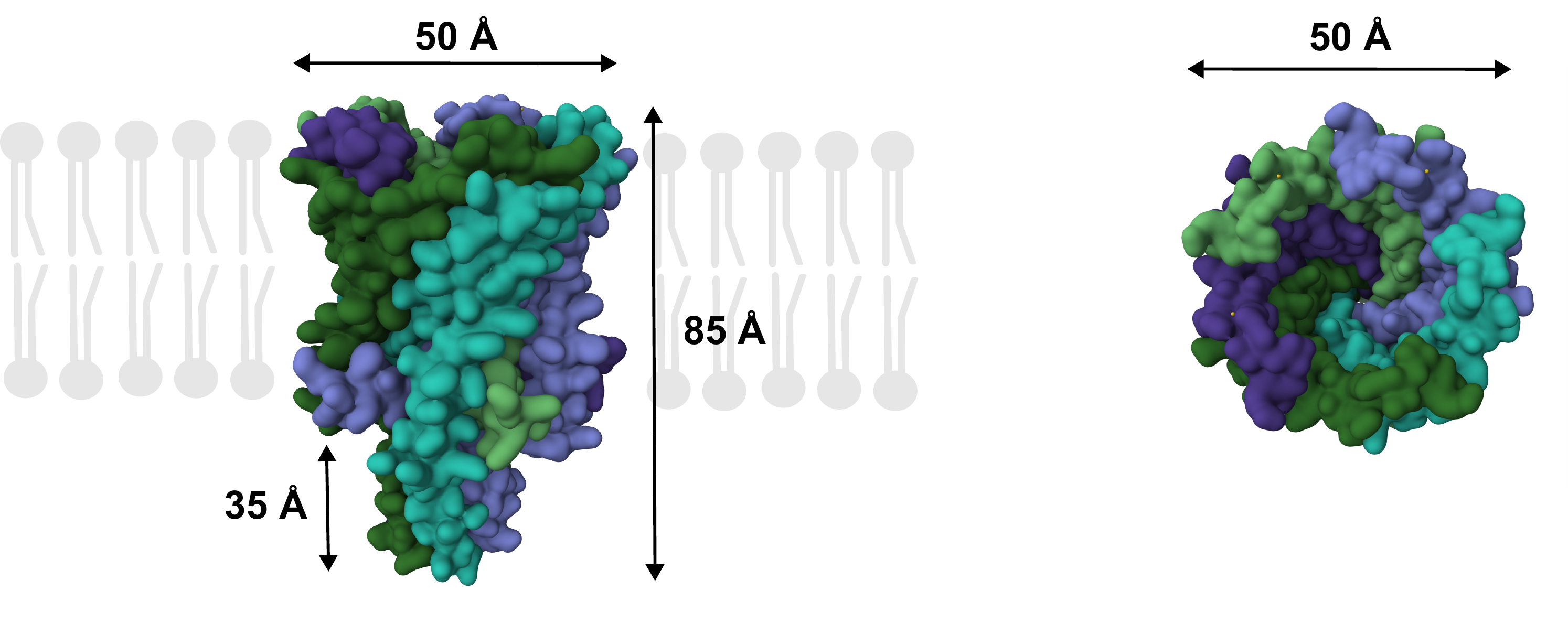}

\caption{Diagram of the closed state of the protein MscL with dimensions in Angstroms
labelled. Dimensions taken from the crystal structure. \cite{chang1998structure} The length of the protein is 85 \AA\ and the C-terminal part of the protrusion is roughly 35 \AA\ in length. In the crystal structure the overall protrusion length of the protein from the cytoplasmic side of the membrane into a cell (bottom) is between 45-50 \AA\ and includes the N-terminus of the protein. MscL molecular surface figures taken from 20AR PDB database
(10.2210/pdb2OAR/pdb). \cite{rose2010rcsb}}
\label{MscLAllDimensions}
\end{figure}

\begin{figure}
\centering
\includegraphics[width=0.5\textwidth]{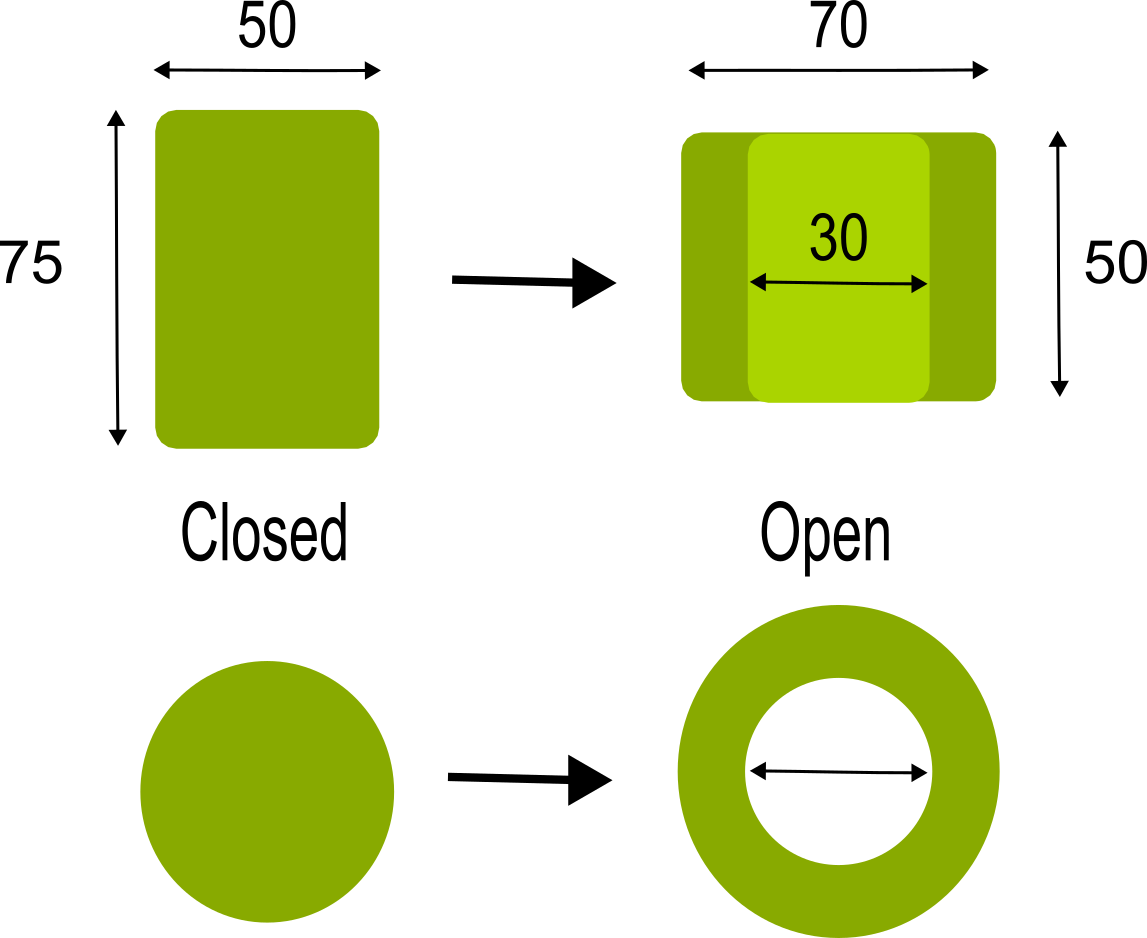}

\caption{Simplified dimensions of the protein MscL, before and after gating assuming retraction of cytoplasmic C-terminal domain.}
\label{MscLGating}
\end{figure}

A simplified approach treats the closed state of MscL as a uniform cylinder, see Figure \ref{MscLGating}, for which the $R_g$ is given by,  
\begin{equation}
R_g^2=(R_0^2+R_1^2)/2+(L^2/12), \label{eq_cyl_Rg}
\end{equation}
where $L$ is the length of the cylinder and $R_1$ is the outer radius of the protein and $R_0$ is the radius of the pore ($R_0=0$ for closed state). 

The change in radius of gyration on gating can be estimated from current open channel models, in which the opening of a 30 \AA\ diameter pore is accompanied by  an increase in protein diameter from 50 \AA{}  to over 70 \AA{} and a retraction of the C-terminus protrusion into the transmembrane region on gating.\cite{perozo2002open}  Taking the literature dimensions for the closed pore from Figure \ref{MscLAllDimensions} and for the open pore from \cite{perozo2002open} , Equation \ref{eq_cyl_Rg} was used to calculate a range of cylinder radius of gyration values dependent on the size of the pore and length of the cylinder, shown in Table \ref{RgforL}.

\begin{table}
    \centering
    \begin{tabular}{ccc}
    \toprule
        & $R_0 = 0, R_1 = 25$ & $R_0 = 15, R_1 = 35$\\ \midrule
       $L = 50$ & 22.8 & 30.6\\
       $L = 55$  & 23.8 & 31.3\\
       $L = 60$ & 24.7 & 32.0\\
       $L = 65$ & 25.8 & 32.8\\
       $L = 70$  & 26.8 & 33.7 \\
       $L = 75$ & \textbf{28.0 }& 34.6 \\
       $L = 80$  & 29.1 & 35.5\\
       $L = 85$  & 30.2 & 36.4\\
       \bottomrule
    \end{tabular}
    \caption{Calculated radius of gyration values for MscL dependent of protein length, $L$, Protein pore radius $R_0$ and Protein radius $R_1$. All values have unit Å. }
   
    \label{RgforL}
\end{table}

In Figure \ref{MscLGating} the simplistic hollow cylinder models are shown with the estimated dimensions, which differ from those shown in Figure \ref{MscLAllDimensions} as there is a much lower mass density in the C-terminal protrusion portion of the protein. With assumed dimensions for the closed state of $L=75$, we use equation \ref{eq_cyl_Rg}  to calculate  $R_g$ = 28 \AA{} for the closed state, in good agreement with the crystal structure derived radius of gyration.

Using these dimensions the radius of gyration on gating changes from 28 \AA{} to between 30.6 \AA \ and 34.6 \AA, dependent on whether there is a total retraction of the C-terminus region into the transmembrane portion or if it remains completely extended.  Note that a larger change in radius of gyration would occur if the C-terminus protrusion does not retract, as some groups have previously suggested. \cite{bavi2017structural}

\subsection{SANS of Lyso-PC and pexiganan interaction with MscL containing vesicles}

Small angle neutron scattering of MscL containing vesicles was measured before and after the addition of the antibacterial molecules lyso-PC and pexiganan in an attempt to observe a conformational change in the protein that would confirm that the protein had been expressed into the vesicles in active form. 

\begin{figure}
\centering

\includegraphics[width=0.7\textwidth]{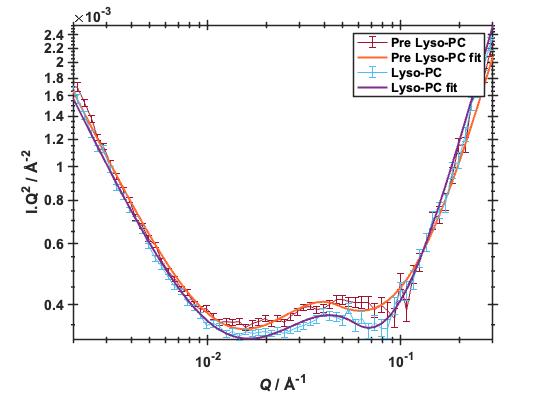}

\includegraphics[width=0.7\textwidth]{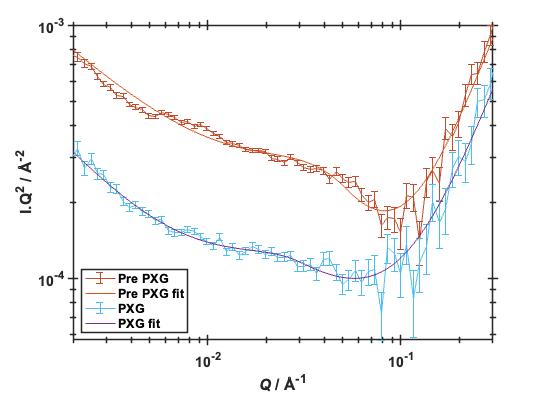}

\caption{Small angle neutron scattering (SANS) of MscL protein containing vesicles before and after the addition of 4~$\mu$M Pexiganan and 4~$\mu$M Lyso-PC measured in D$_2$O 20~mM HEPES buffer pD~7.4. 2-level Guinier-Porod model shown in as solid line, experimental SANS data is displayed as error bars.}
\label{LysoPCandpexiganan}
\end{figure}

A concentrated stock of lyso-PC (5~$\mu$L, 400~$\mu$M), a single-tailed lipid used to permanently gate MscL in previous studies,\cite{martinac1990mechanosensitive, perozo2002open,Strutt21} was added to a solution of MscL containing vesicles (500~$\mu$L) to a final lyso-PC concentration of 4~$\mu$M. The vesicles were characterised by small angle neutron scattering before and after lyso-PC addition. The Kratky plot shown in the upper panel of Figure~\ref{LysoPCandpexiganan} shows that addition of lyso-PC induces a small change in the scattering from the MscL containing vesicles. The best fit of this data to a Guinier-Porod model is consistent with  the radius of gyration  corresponding to the GP1 region associated with the protein increasing from 26$\pm$3~\AA{} to 30$\pm$1~\AA{} (Table~\ref{GPtable2}). As we showed above in section~\ref{MscL_Rg}, on gating the radius of gyration of an individual MscL channel can be estimated to increase from from 28 to between 30.6 and 34.6~\AA{}. This small increase in $R_g$ accompanying the addition of lyso-PC could be a signature of an increase in the fraction of MscL channels that are gated open.

\begin{figure}
\centering
\includegraphics[scale=0.18]{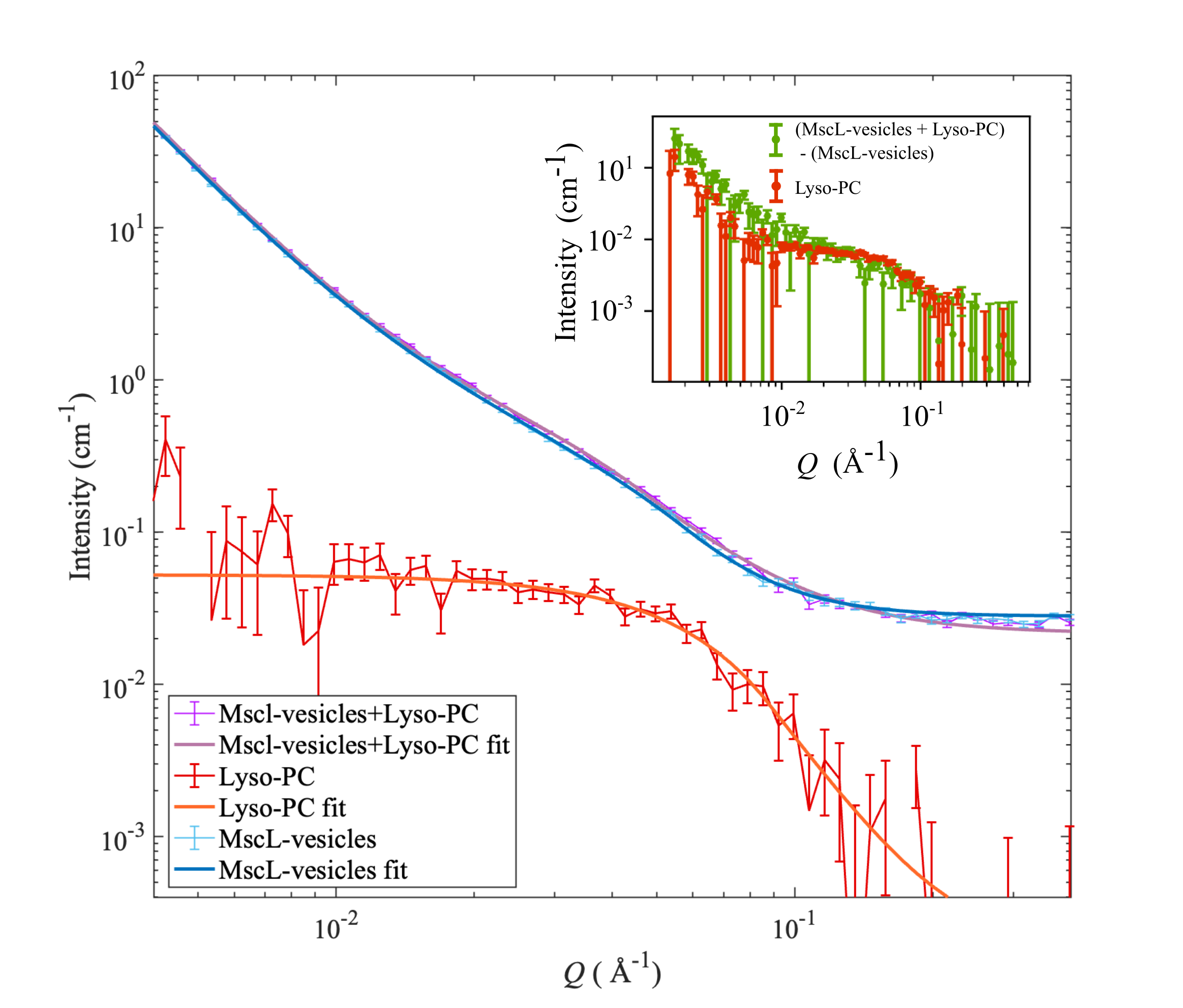}

\caption{SANS of MscL containing vesicles measured in D$_2$O 20 mM HEPES buffer at pD 7.4 before (blue errorbars) and after (pink errorbars) the addition of 4$\mu$M Lyso-PC. The corresponding 2-level Guinier-Porod fits are shown as light blue and purple lines, respectively. Also shown (red errorbars) is the SANS measured from the 4$\mu$M Lyso-PC in the same buffer and the corresponding fit to a spherical micelle model (orange line). Inset figure shows a subtracted dataset showing the difference in scattering of the MscL proteoliposomes before and after the addition of lyso-PC (green errorbars) overlaid with the scattering from lyso-PC micelles (red errorbars).}
\label{lysopcoverlay}
\end{figure}

Figure \ref{lysopcoverlay} shows the SANS measured from the 4~$\mu$M solution of lyso-PC in D$_2$O buffer (red data points) and the corresponding fit to a Guinier-Porod model (best fit radius=27.7Å) plotted on the same scale as the scattering measured from the MscL containing vesicles before (black data points) and after (blue data points) the addition of the lyso-PC. 
In the inset figure of Figure \ref{lysopcoverlay}, we have plotted the data set of lyso-PC interacted MscL containing vesicles with the 'prior to lyso-pc interaction data' subtracted, and displayed this with the SANS from lyso-PC micelles. From this we show that scattering from lyso-PC micelles alone cannot account for the change observed following addition of lyso-PC to the MscL-containing vesicles. The overlap in the scattering means that we cannot make strong conclusions on change in radius of gyration from this data set, however the change in vesicle scattering suggests a conformational change has occurred and that our proteins display lyso-PC gating activity, as demonstrated by other groups. \cite{grage2011bilayer}

The effect of pexiganan was tested at a concentration of 4~$\mu$M, a concentration slightly above that of its minimum inhibitory concentration (MIC), to allow for a direct comparison to the effect of lyso-PC measured at the same concentration. A solution of pexiganan in D$_2$O buffer (5 ~$\mu$L at 400~$\mu$M) was  added to  500~$\mu$L of the MscL containing vesicles in D$_2$O buffer. The Kratky plot shown in the lower panel of Figure \ref{LysoPCandpexiganan} shows the change in scattering profile  induced by the addition of pexiganan. Addition of pexiganan, results in  a 62.5\% decrease in the scattered intensity, $I(Q=2\times 10^{-3}$~\AA$^{-1})$ as a result of a decrease in the stability of the MscL containing vesicles, which results leads to some creaming of lipid/protein from the dispersion, removing it from the beam path. 

The best-fit to a 2-level Guinier-Porod model for the SANS measured before and after the addition of pexiganan  is provided by the parameters given in the lower panel of Table~\ref{GPtable2}. In this case the $R_g$ increases from 29$\pm$4~\AA{}  to 36$\pm$3~\AA{}. As this is comparable to the change we predicted in section~\ref{MscL_Rg}, using a simple cylinder model for MscL  to accompany gating, we infer that not only is the MscL channel active in the vesicles, but that a greater proportion of the channels are gated by the interaction of pexiganan than lyso-PC at 4~$\mu$M.

\begin{table*}
\centering
\caption{Best fit parameters of 2-level Guinier-Porod model to MscL protein containing vesicles before and after the addition of 4 ~$\mu$M lyso-PC or 4 ~$\mu$M pexiganan in D$_2$O buffer, the resulting models are displayed in Figure \ref{LysoPCandpexiganan}. }
\begin{tabular}{|c|l|l|}
\hline
Parameter & Before addition &	After addition of 4 ~$\mu$M lyso-PC\\
\hline
$GP1$ & &\\
\hline
$s$	& 0.96 $\pm$ 0.3 &	0.78 $\pm$ 0.04 \\
$G$	& $(1.33 \pm$ 0.17)$\times10^{-2}$ & $(3.00 \pm$ 0.46) $\times10^{-2}$	\\
$R_g$(Å)	& 26.1 $\pm$ 2.6 &	29.5 $\pm$ 1.2 \\
$d$	& 3.2 $\pm$ 0.3 & 2.47 $\pm$ 0.05 \\
\hline
$GP2$ & &\\
\hline
$s$	&	0 (fixed) & 0 (fixed)\\
$G$	&	(3.9$ \pm $0.1)  $\times10^{-12}$ & (3.5 $\pm$ 0.2)$\times10^{-12}$ \\
$R_g$(Å) & 10$^6$ (fixed) & 10$^6$ (fixed) \\
$d$	&  3.14 $\pm$ 0.08 & 3.12 $\pm$ 0.06\\
\hline
$\chi^2$	& 72.1 &  120\\
Normalised $\chi^2$	& 1.08 & 1.8 \\
\hline
Parameter & Before addition &	After addition of 4~$\mu$M pexiganan\\
\hline
$GP1$ & &\\
\hline
$s$	& 1.21 $\pm$ 0.03 &	0.86 $\pm$ 0.09\\
$G$	& $(2.7 \pm$ 0.2)$\times 10^{-3}$ &	(5.9 $ \pm$ 5.5)$\times 10^{-3}$ \\
$R_g$(Å)	& 29.0 $\pm$ 4.0 &	35.8 $\pm$ 3.3 \\
$d$	& 3.5 (fixed) & 3.5 (fixed)  \\
\hline
$GP2$ & &\\
\hline
$s$	&	0 (fixed) & 0 (fixed)\\
$G$	&	(2.22 $\pm$ 0.29)$\times10^{-10}$ & (1.68 $\pm$ 0.16)$\times10^{-10}$  \\
$R_g$(\AA) & 10$^6$ (fixed) & 10$^6$ (fixed) \\
$d$	&  2.51 $\pm$ 0.02 & 2.59 $\pm$ 0.01\\
\hline
$\chi^2$	& 200 &  93\\
Normalised $\chi^2$	& 1.3 &  1.1\\
\hline
\end{tabular}
\label{GPtable2}
\end{table*}

\section{Reflectivity model implemented in Python using RefNX}\label{listing}

\lstset{breaklines=true}
\begin{lstlisting}[language=Python, caption=Python code for fitting polarized neutron reflectivity data measured from the tethered bilayer in three sub-phase contrasts. The full set of Jupyter notebooks used for the analysis of PNR can be found in the data repository for the manuscript~\cite{Ayscough_Tethered_Lipid_Bilayers_2025}.]
#!/usr/bin/env python
# coding: utf-8

# In[48]:


#@title Imports
# use matplotlib for plotting
get_ipython().run_line_magic('matplotlib', 'inline')
import matplotlib.pyplot as plt
import numpy as np
import os.path
import pandas as pd
import numpy as np
import math
from scipy import special

# from ipysheet import sheet, column
# import qgrid
import pandas as pd

# from google.colab import output
# output.enable_custom_widget_manager()

import refnx, scipy

# the analysis module contains the curvefitting engine
from refnx.analysis import CurveFitter, Objective, Parameter, GlobalObjective, process_chain

# the reflect module contains functionality relevant to reflectometry
from refnx.reflect import SLD, ReflectModel, Structure, LipidLeaflet, Slab

# the ReflectDataset object will contain the data
from refnx.dataset import ReflectDataset

get_ipython().run_line_magic('matplotlib', 'inline')
plt.style.use('classic')
plt.rcParams['figure.figsize'] = [10, 5]
plt.rcParams['axes.linewidth'] = 2
plt.rcParams['axes.facecolor'] = 'white'
plt.rcParams['mathtext.default'] = 'regular'

# set tick width
plt.rcParams['xtick.major.size'] = 10
plt.rcParams['xtick.major.width'] = 3
plt.rcParams['ytick.major.size'] = 10
plt.rcParams['ytick.major.width'] = 3


# In[49]:


conditions = ['bilayer', '1p6', '3p2']

condition = conditions[0]

if condition == '1p6':
    pth = '/mnt/ceph/home/ms9743/analysis/TetherBayes2022/JAN21_Rascal2020_bil_1p6PXG/dataFiles/'

    data_d2o_up = ReflectDataset(pth + 'POLLREFfinalIvsQ_26854_26855_26856_IvsQ_26854_1_IvsQ_26855_1_IvsQ_26856_1.dat.txt')
    data_d2o_up.name = "D$_2$O up"

    data_d2o_down = ReflectDataset(pth + 'POLLREFfinalIvsQ_26854_26855_26856_IvsQ_26854_2_IvsQ_26855_2_IvsQ_26856_2.dat.txt')
    data_d2o_down.name = "D$_2$O down"

    data_gmw_up = ReflectDataset(pth + 'POLLREFfinalIvsQ_26860_26861_26862_IvsQ_26860_1_IvsQ_26861_1_IvsQ_26862_1.dat.txt')
    data_gmw_up.name = "GMW up"

    data_gmw_down = ReflectDataset(pth + 'POLLREFfinalIvsQ_26860_26861_26862_IvsQ_26860_2_IvsQ_26861_2_IvsQ_26862_2.dat.txt')
    data_gmw_down.name = "GMW down"

    data_h2o_up = ReflectDataset(pth + 'POLLREFfinalIvsQ_26863_26864_26865_IvsQ_26863_1_IvsQ_26864_1_IvsQ_26865_1.dat.txt')
    data_h2o_up.name = "H$_2$O up"

    data_h2o_down = ReflectDataset(pth + 'POLLREFfinalIvsQ_26863_26864_26865_IvsQ_26863_2_IvsQ_26864_2_IvsQ_26865_2.dat.txt')
    data_h2o_down.name = "H$_2$O down"
    
elif condition == '3p2':
    pth = '/mnt/ceph/home/ms9743/analysis/TetherBayes2022/JAN21_Rascal21_bil_3p2PXG/dataFiles/'

    data_d2o_up = ReflectDataset(pth + 'POLLREFfinalIvsQ_26875_26876_26877_IvsQ_26875_1_IvsQ_26876_1_IvsQ_26877_1.dat.txt')
    data_d2o_up.name = "D$_2$O up"

    data_d2o_down = ReflectDataset(pth + 'POLLREFfinalIvsQ_26875_26876_26877_IvsQ_26875_2_IvsQ_26876_2_IvsQ_26877_2.dat.txt')
    data_d2o_down.name = "D$_2$O down"

    data_gmw_up = ReflectDataset(pth + 'POLLREFfinalIvsQ_26878_26879_26880_IvsQ_26878_1_IvsQ_26879_1_IvsQ_26880_1.dat.txt')
    data_gmw_up.name = "GMW up"

    data_gmw_down = ReflectDataset(pth + 'POLLREFfinalIvsQ_26878_26879_26880_IvsQ_26878_2_IvsQ_26879_2_IvsQ_26880_2.dat.txt')
    data_gmw_down.name = "GMW down"

    data_h2o_up = ReflectDataset(pth + 'POLLREFfinalIvsQ_26881_26882_26883_IvsQ_26881_1_IvsQ_26882_1_IvsQ_26883_1.dat.txt')
    data_h2o_up.name = "H$_2$O up"

    data_h2o_down = ReflectDataset(pth + 'POLLREFfinalIvsQ_26881_26882_26883_IvsQ_26881_2_IvsQ_26882_2_IvsQ_26883_2.dat.txt')
    data_h2o_down.name = "D$_2$O down"
    
elif condition == 'bilayer':
    pth = '/home/ms9743/analysis/TetherBayes2022/DEC21_Rascal2020_bilayer/dataFiles/'

    data_d2o_up = ReflectDataset(pth + 'POLLREFfinalIvsQ_26838_26839_26840_IvsQ_26838_1_IvsQ_26839_1_IvsQ_26840_1.dat')
    data_d2o_up.name = "D$_2$O up"

    data_d2o_down = ReflectDataset(pth + 'POLLREFfinalIvsQ_26838_26839_26840_IvsQ_26838_2_IvsQ_26839_2_IvsQ_26840_2.dat')
    data_d2o_down.name = "D$_2$O down"

    data_gmw_up = ReflectDataset(pth + 'IvsQ_26841_26842_26843_IvsQ_26841_1_IvsQ_26842_1_IvsQ_26843_1.dat.txt')
    data_gmw_up.name = "GMW up"

    data_gmw_down = ReflectDataset(pth + 'IvsQ_26841_26842_26843_IvsQ_26841_2_IvsQ_26842_2_IvsQ_26843_2.dat.txt')
    data_gmw_down.name = "GMW down"

    data_h2o_up = ReflectDataset(pth + 'IvsQ_26844_26845_26846_IvsQ_26844_1_IvsQ_26845_1_IvsQ_26846_1.dat.txt')
    data_h2o_up.name = "H$_2$O up" 

    data_h2o_down = ReflectDataset(pth + 'IvsQ_26844_26845_26846_IvsQ_26844_2_IvsQ_26845_2_IvsQ_26846_2.dat.txt')
    data_h2o_down.name = "H$_2$O down"


# In[50]:


# Define some SLDs
si_sld = SLD(2.07 + 0j)
sio2_sld = SLD(3.47 + 0j)

# au_sld = SLD(4.2 + 0j)
# au_sld.real.setp(vary=False) #, bounds=(4.2, 4.7))

PEG_sld = SLD(0.622 + 0j)
PDP_sld = SLD(1.01 + 0j)

# the following represent the solvent contrasts used in the experiment
d2o = SLD(6.36 + 0j)
h2o = SLD(-0.56 + 0j)
gmw = SLD(4.5 + 0j)

# We want the `real` attribute parameter to vary in the analysis, and we want to apply
# uniform bounds. The `setp` method of a Parameter is a way of changing many aspects of
# Parameter behaviour at once.
d2o.real.setp(vary=True, bounds=(6.1, 6.36))
d2o.real.name='d2o SLD'

h2o.real.setp(vary=True, bounds=(-0.56, 0.0))
h2o.real.name='h2o SLD'

gmw.real.setp(vary=True, bounds=(4.1, 4.7))
gmw.real.name='gmw SLD'


# In[51]:


#@title Parameter


# define model parameters [value, min, max, vary?]
data = {
        'PDP_thick'    : [6, 5, 15, True],
        'PDP_hydration': [0.25, 0.2, 0.6, True],
        'PEG_thick'    : [73, 40, 100, True],
        'PEG_hydration': [0.53, 0.2, 1, True],
        'PEG_rough'    : [11, 5, 25, True],
        'lipid_APM'    : [66, 50, 80, True],
        'WPLH'         : [14.8, 5, 30, True],
        'PCoverage'    : [0.125, 0, 0.3, True],
        'protrus_thick': [50, 0, 65, False],
        'lipid_APMT'   : [80, 40, 100, True],
        'WPLT'         : [8.6, 0, 20, True]
        }

    
# Load best fit parameters from tetjer only fit:  
df = pd.read_pickle('tether_only_params.pkl') #pd.DataFrame.from_dict(data, orient='index', columns=['Value', 'Min', 'Max', 'Vary'])

for par in data:
  locals()[par] = par
  locals()[par] = Parameter(data[par][0], par, 
                            vary=data[par][3], 
                            bounds=(data[par][1], data[par][2]))


# generate parameters from df
for par in df.T:
  locals()[par] = df.loc[par].name
  locals()[par] = Parameter(df.loc[par].value, df.loc[par].name, 
                            vary=True, 
                            bounds=(df.loc[par].bounds.lb, df.loc[par].bounds.ub))

# Fix hard layer parameters from thether only fit
oxide_thick.setp(vary=False)
substr_rough.setp(vary=False)
py_thick.setp(vary=False)
py_rough.setp(vary=False)
py_SLD_up.setp(vary=False)
py_SLD_down.setp(vary=False)
au_sld.setp(vary=False)
au_thick.setp(vary=False)
au_rough.setp(vary=False)


# bilayer coverage  
bilayer_coverage = Parameter(0.8, 'bilayer coverage')
bilayer_coverage.setp(vary=True, bounds=(0, 1))

df


# In[52]:


protrus_thick.setp(vary=True)


# In[53]:


#@title Bilayer model definition
def bilayer(solvent):
  # Calculations...
  #Define all the neutron b's
  bc = 0.6646e-4	#Carbon
  bo = 0.5843e-4	#Oxygen
  bh = -0.3739e-4	#Hydrogen
  bp = 0.513e-4	#Phosphorus
  bn = 0.936e-4	#Nitrogen
  bd = 0.6671e-4	#Deuterium
  bs = 2.847e-4 #Sulphur

  #Formulae of the molecule sections...
  CH = (1*bc) + (1*bh)
  CH2 = (1*bc) + (2*bh)
  CH3 = (1*bc) + (3*bh)
  D2O = (2*bd) + (1*bo)
  H2O = (2*bh) + (1*bo)
#   print(D2O)
  # PDP_b = (2*bn) + (7*bc) + (9*bh) + (2*bs) + (1*bo);
  # PEG_unit = (2*CH2)+ (1*bo);
  PEG_sld = 6.22e-7
  PDP_sld = 1.01e-6

  # Calculate mole fraction of D2O from the bulk SLD..
  d2o_molfr = (1/D2O-H2O)*((solvent.real.value*27.64)-H2O)
  wMol = (d2o_molfr * D2O) + ((1-d2o_molfr)*H2O)

  # sum b's of all the different fragments
  sum_b_tails = -0.00026668800000000006 # (28*CH2) + (2*CH) + (2*CH3) + (WPLT * wMol)
  sum_popc_heads = 0.0006007780000000002 # (8*bo) + (1*bp) + (1*bn) + (2*bc) + (4*CH2) + (3*CH3) + (1*CH)
  sum_popg_heads = 0.0009230619999999998 # (10*bo) + (1*bp) + (2*bc) + (4*CH2) + (2*CH) + (2*bh)
  sum_b_heads = (1/4)*(3*(sum_popc_heads) + (sum_popg_heads)) + (WPLH * wMol)
  sum_b_headsT = (1/4)*(3*(sum_popc_heads) + (sum_popg_heads)) + (WPLH * wMol)
  sum_mscl = (1990*bc) +(455*bn) + (433*bo) + (20*bs) + (((3157.5-(594.5*0.7))*bh)) + (d2o_molfr*(594.5*0.7)*bd) + ((1-d2o_molfr)*(594.5*0.7)*bh)

  #Cytoplasmic section
  sum_mscl_cterminal = (910*bc) +(280*bn) + (296*bo) + ((1467-(387*0.9))*bh) + (d2o_molfr*(387*0.9)*bd) + ((1-d2o_molfr)*(387*0.9)*bh)
  sum_mscl_nterminal = (315*bc) +(75*bn) + (81*bo) + (5*bs) + ((467-(102*0.9))*bh) + (d2o_molfr*(102*0.9)*bd) + ((1-d2o_molfr)*(102*0.9)*bh)
  sum_mscl_histag = (180*bc) +(90*bn) + (31*bo) + ((197-(47*0.9))*bh) + (d2o_molfr*(47*0.9)*bd) + ((1-d2o_molfr)*(47*0.9)*bh)
  sum_mscl_cyto = sum_mscl_cterminal + sum_mscl_nterminal + sum_mscl_histag
    

  #volumes of each fragment
  vol_w = 29.7
  volume_tails = 944 #+ (WPLT * vol_w)
  vfsolv_tails = 1-bilayer_coverage #(WPLT * vol_w)/volume_tails
    
  volume_heads = ((1/4)*((3*319)+(257)))+ (WPLH * vol_w) #R.Armen and J.Pan
  vfsolv_heads = (WPLH * vol_w)/volume_heads
    
  volume_headsT = ((1/4)*((3*319)+(257)))+ (WPLH * vol_w)
  vfsolv_headsT = (WPLH * vol_w)/volume_headsT
    
  volume_mscl_cterminus = 25979.5
  volume_mscl_nterminus = 8525
  volume_histag= 4719
  volume_mscl_cyto = volume_mscl_cterminus + volume_mscl_nterminus + volume_histag
  volume_mscl = 53987+6225.5

  LTailThick = volume_tails / lipid_APM
  LHeadThick = volume_heads / lipid_APM
  LTailThickT = volume_tails / lipid_APMT
  LHeadThickT = volume_headsT / lipid_APMT

  Rho_heads = sum_b_heads / volume_heads
  Rho_headsT = sum_b_headsT / volume_headsT
  Rho_tails = sum_b_tails / volume_tails
  Rho_mscl = sum_mscl / volume_mscl

  Rho_mscl_cyto = sum_mscl_cyto / volume_mscl_cyto
  Rho_h_m = ((1-PCoverage)*Rho_heads) + ((PCoverage)*Rho_mscl)
  Rho_t_m = ((1-PCoverage)*Rho_tails) + ((PCoverage)*Rho_mscl)
  Rho_h_mT = ((1-PCoverage)*Rho_headsT) + ((PCoverage)*Rho_mscl)
#   Ccoverage = ((volume_mscl_cyto*(2*(LHeadThick+LTailThick)))/(protrus_thick*volume_mscl))*(PCoverage)
  Ccoverage = ((volume_mscl_cyto*((LHeadThick+LTailThick+LHeadThickT+LTailThickT)))/(protrus_thick*volume_mscl))*(PCoverage)
  #Lcoverage = ((volume_mscl_periplasmic*(2*(HeadThick+TailThick)))/(mscl_prot_thick_t*volume_mscl))*(PCoverage)
  Rho_mscl_prot = (Ccoverage)*Rho_mscl_cyto + ((1-Ccoverage)*(solvent.real.value))
  


# make structue out of this:
  
  inner_head_l = Slab(LHeadThick, Rho_h_m, PEG_rough , name='Inner HG', vfsolv=0, interface=None)
  inner_tail_l = Slab(LTailThick, Rho_t_m, PEG_rough , name='Inner Tail', vfsolv=1-bilayer_coverage, interface=None) 
  outer_head_l = Slab(LHeadThickT, Rho_h_mT, PEG_rough , name='Outer HG', vfsolv=0, interface=None)
  outer_tail_l = Slab(LTailThickT, Rho_t_m, PEG_rough , name='Outer Tail', vfsolv=1-bilayer_coverage, interface=None) 

#   mscl_prot_l = Slab(protrus_thick, Rho_mscl_prot, PEG_rough , name='mscl_prot_layer', vfsolv=0, interface=None)
  mscl_prot_l = Slab(protrus_thick, Rho_mscl_cyto, PEG_rough , name='MscL', vfsolv=(1-Ccoverage), interface=None)

  s_bilayer = inner_head_l | inner_tail_l | outer_tail_l | outer_head_l | mscl_prot_l# | solvent(0, solv_roughness)

  total_thickness = LTailThick + LHeadThick + LTailThickT + LHeadThickT
    
  return s_bilayer, total_thickness, PCoverage, vfsolv_heads, vfsolv_tails, vfsolv_headsT, Ccoverage


# In[54]:


# define layers

solv_roughness = Parameter(3, 'bilayer/solvent roughness')
solv_roughness.setp(vary=True, bounds=(2, 15))

oxide_l = Slab(oxide_thick, sio2_sld, substr_rough , name='oxide_layer', vfsolv=0, interface=None)
py_up_l = Slab(py_thick, py_SLD_up, py_rough , name='py_up_layer', vfsolv=0, interface=None)
py_down_l = Slab(py_thick, py_SLD_down, py_rough , name='py_down_layer', vfsolv=0, interface=None)
au_l = Slab(au_thick, au_sld, au_rough , name='Au', vfsolv=0, interface=None)
PDP_l = Slab(PDP_thick, PDP_sld, au_rough , name='PDP', vfsolv=PDP_hydration, interface=None)
PEG_l = Slab(PEG_thick, PEG_sld, PEG_rough , name='PEG', vfsolv=PEG_hydration, interface=None)

s_d2o_up = si_sld | oxide_l | py_up_l | au_l | PDP_l | PEG_l | bilayer(d2o)[0] | d2o(0, solv_roughness)
s_gmw_up = si_sld | oxide_l | py_up_l | au_l | PDP_l | PEG_l | bilayer(gmw)[0] | gmw(0, solv_roughness) 
s_h2o_up = si_sld | oxide_l | py_up_l | au_l | PDP_l | PEG_l | bilayer(h2o)[0] | h2o(0, solv_roughness)

s_d2o_down = si_sld | oxide_l | py_down_l | au_l | PDP_l | PEG_l | bilayer(d2o)[0] | d2o(0, solv_roughness)
s_gmw_down = si_sld | oxide_l | py_down_l | au_l | PDP_l | PEG_l | bilayer(gmw)[0] | gmw(0, solv_roughness) 
s_h2o_down = si_sld | oxide_l | py_down_l | au_l | PDP_l | PEG_l | bilayer(h2o)[0] | h2o(0, solv_roughness)


# In[55]:


model_scale = Parameter(1,'common scale')
model_scale.setp(vary=True, bounds=(0.9,1.1))

# Define models
qres=4.0
model_d2o_up = ReflectModel(s_d2o_up, scale=model_scale, dq=qres)
model_d2o_down = ReflectModel(s_d2o_down, scale=model_scale, dq=qres)

model_gmw_up = ReflectModel(s_gmw_up, scale=model_scale, dq=qres)
model_gmw_down = ReflectModel(s_gmw_down, scale=model_scale, dq=qres)

model_h2o_up = ReflectModel(s_h2o_up, scale=model_scale, dq=qres)
model_h2o_down = ReflectModel(s_h2o_down, scale=model_scale, dq=qres)

model_d2o_up.scale.setp(vary=True, bounds=(0.9, 1.1))

# Backgrounds
model_d2o_up.bkg.setp(3e-6, vary=True, bounds=(1e-7, 1e-5))
model_gmw_up.bkg.setp(3e-6, vary=True, bounds=(1e-7, 1e-5))
model_h2o_up.bkg.setp(3e-6, vary=True, bounds=(1e-7, 1e-5))

model_d2o_down.bkg.setp(3e-6, vary=True, bounds=(1e-7, 1e-5))
model_gmw_down.bkg.setp(3e-6, vary=True, bounds=(1e-7, 1e-5))
model_h2o_down.bkg.setp(3e-6, vary=True, bounds=(1e-7, 1e-5))


aux_pars = [WPLH, lipid_APMT, lipid_APM, bilayer_coverage, protrus_thick, PCoverage, model_scale]
objective_d2o_up = Objective(model_d2o_up, data_d2o_up, auxiliary_params=aux_pars)
objective_d2o_down = Objective(model_d2o_down, data_d2o_down, auxiliary_params=aux_pars)

objective_gmw_up = Objective(model_gmw_up, data_gmw_up, auxiliary_params=aux_pars)
objective_gmw_down = Objective(model_gmw_down, data_gmw_down, auxiliary_params=aux_pars)

objective_h2o_up = Objective(model_h2o_up, data_h2o_up, auxiliary_params=aux_pars)
objective_h2o_down = Objective(model_h2o_down, data_h2o_down, auxiliary_params=aux_pars)

global_objective = GlobalObjective([objective_d2o_up, objective_gmw_up, objective_h2o_up,
                                   objective_d2o_down, objective_gmw_down, objective_h2o_down])

# create the fit instance
fitter = CurveFitter(global_objective)
np.random.seed(1)
fitter.initialise('jitter')

# fitter.fit('differential_evolution')
# fitter.fit('least_squares')

# In[60]:
fitter.sample(400, random_state=1)
fitter.sampler.reset()
fitter.sample(50, nthin=100, random_state=1)
\end{lstlisting}

\bibliography{Ayscough25}
\bibliographystyle{elsarticle-num}